\documentclass[twocolumn,aps,prc,superscriptaddress,showpacs]{revtex4-1}

\usepackage{verbatim}
\usepackage{graphicx}
\usepackage{epsfig}
\usepackage{amsfonts}
\usepackage{amsmath}
\usepackage{amssymb}
\usepackage{bm}
\usepackage{hyperref}
\usepackage{color}
\usepackage{float}
\usepackage{siunitx}
\usepackage{multirow}
\usepackage{pgf}
\usepackage{subcaption}

\begin{document}

\title{Estimate of the location of the neutron drip line for calcium isotopes from an exact Hamiltonian with continuum pair correlations}

\author{A.C. Dassie}
\affiliation{Physics Institute of Rosario (CONICET-UNR), 
             Esmeralda y Ocampo, S2000EZP Rosario, Argentina}
\affiliation{Department of Physics FCEIA (UNR),
             Av. Pellegrini 250, S2000BTP Rosario, Argentina}
\affiliation{Institute of Nuclear Studies and Ionizing Radiations (UNR), 
		    	Riobamba y Berutti, S2000EKA Rosario, Argentina.}             
		    	
\author{R.M. Id Betan}
\affiliation{Physics Institute of Rosario (CONICET-UNR), 
             Esmeralda y Ocampo, S2000EZP Rosario, Argentina}
\affiliation{Department of Physics FCEIA (UNR),
             Av. Pellegrini 250, S2000BTP Rosario, Argentina}
\affiliation{Institute of Nuclear Studies and Ionizing Radiations (UNR), 
		    	Riobamba y Berutti, S2000EKA Rosario, Argentina.}             

%date{\today}

\begin{abstract}
  \begin{description} 
    \item[Background] The eastern region of the calcium isotope chain of the nuclei chart is, nowadays, of great activity. The experimental assessment of the limit of stability is of interest to confirm or improve microscopic theoretical models. 
    \item[Purpose] The goal of this work is to provide the drip line of the calcium isotopes from the exact solution of the pairing Hamiltonian which incorporate explicitly the correlations with the continuum spectrum of energy.
    \item[Method] The modified Richardson equations, which include correlations with the continuum spectrum of energy modeled by the continuum single particle level density, is used to solve the many-body system. Three models are used, two isospin independent models with core $^{40}$Ca and $^{48}$Ca, and one isospin dependent model.
    \item[Results] One and two-neutron separation energies and occupation probabilities for bound and continuum states are calculated from the solution of the Richardson equations.
    \item[Conclusions] The one particle drip line is found at the nucleus $^{57}$Ca, while the two neutron drip line is found at the nucleus $^{60}$Ca from the isospin independent model and at $^{66}$Ca from the isospin dependent one.
  \end{description}
\end{abstract}

\pacs{04.20.Jb,21.10.Dr,21.60.-n,24.10.Cn,27.50.+e}
%\keywords{Suggested keywords}%Use showkeys class option if keyword display desired
%
%% 04.20.Jb Exact solutions
%% 21.10.Gv Nucleon distributions and halo features
%% 21.10.Ma Level density
%% 21.10.Pc Single-particle levels and strength functions
%% 21.10.-k 	Properties of nuclei; nuclear energy levels
%% 21.10.Dr 	Binding energies and masses
%% 21.10.Ma 	Level density
%% 21.30.Fe Forces in hadronic systems and effective interactions
%% 21.60.-n Nuclear structure models and methods
%% 21.60.Cs Shell model
%% 21.60.Ev Collective models
%% 24.10.Cn Many-body theory
%% 27.20.+n Properties of specific nuclei listed by mass ranges: 6 ≤ A ≤ 19
%% 27.30.+t 20 ≤ A ≤ 38
%% 27.40.+z 39 ≤ A ≤ 58
%% 27.50.+e 59 ≤ A ≤ 89
%% 27.80.+w 	190 ≤ A ≤ 219( 	Properties of specific nuclei listed by mass ranges)
                  
\maketitle

%------------------
\section{Introduction} \label{sec.introduction}
The lego-like construction of isotopes for a given atomic nucleus, sooner or later faces the particle continuum. For example, the last observed bound Fluorine is $^{31}$F \cite{2019Ahn}, while the last bound Oxygen is  $^{24}$O \cite{1970Artukh,1985Langevin,1990Mueller,2012Lunderberg}. This simple comparison between two elements which defer only in a single proton, shows the complicated character of drip lines systems, posing a big challenge to nuclear structure models. Interaction \cite{2010Otsuka}, continuum \cite{DavidM.Brink938} and many-body correlations \cite{2013Forssen}, all together collude in this \mbox{kingdom \cite{2012Erler,2020Johnson}.}

The isotopic chain of calcium is currently under scrutiny from both the theoretical and experimental aspects. A handful of nuclei $^{59}$Ca and $^{60}$Ca have been recently observed \cite{2018Tarasov}, they are the heaviest calcium isotopes discovery up to today and both were found to be bound. Their masses are not known yet, the more recent measured atomic mass is that of $^{57}$Ca \cite{2018Michimasa}. The calcium chain is also interesting because it allows the investigation for existence of doubly magic nuclei and the evolution of the charge radius \cite{2013Wienholtz,2013Steppenbeck,2016Gade,2016Ruiz,2019Tanaka,2019Liu,2020Leistenschneider,2020Cortes}.

This paper focuses on the stability limit of the calcium isotopes. We have to wait for updating \cite{2017nupecc} or finishing some facilities to get masses for isotopes of calcium beyond $^{57}$Ca. For example, the Facility for Rare Isotope Beams (FRIB) \cite{2018frib} will measure the key nucleus $^{60}$Ca, recently discovered at RIKEN \cite{2018Tarasov}. Meanwhile, different theoretical approaches are implemented to predict the calcium drip line. Some formalism predicts it as soon as around $^{60}$Ca \cite{2012Hagen,2013Forssen,2018Tichai,2020Soma}, while others predict the drip line at $^{68}$Ca \cite{2014Agbemava,2019Neufcourt}, or even $^{70}$Ca \cite{2012Erler,2018Tarasov,2020Li}. 

Pairing encompasses an important part of the short-range interaction between the neutrons \cite{1964Lane,DavidM.Brink938}. Various approaches have been developed in the last fifty years \cite{2013Zelevinsky} to incorporate pairing in finite nuclei. The Gorkov field theory approach \cite{1958Gorkov,2011Gorkov} properly account for the pairing correlations in many-body systems. Its application to finite nuclei was recently developed \cite{2011Soma,2014Soma} and applied to the calcium isotopes \cite{2013Soma,2020Soma}. Exact results are important in many-body systems, the algebraic Gorkov solution for the separable interaction was given in Ref. \cite{2020IdBetan}, while in this paper we study the calcium chain from the exact solution of the pairing Hamiltonian \cite{Richardson1963,1964Richardson,IdBetan2012}. The correlations with the continuum spectrum of energy is included through the  continuum single particle level density \cite{IdBetan2012P}.

In section \ref{sec.method} we give the theoretical tools used in this paper; with the outline of the method for solving exactly the many-body system with pairing in the continuum for even nuclei, in subsection \ref{sec.methodA}. In subsection \ref{sec.methodB} we relate the calculated magnitudes with the occupation probabilities and the binding energy for even and odd isotopes. In section \ref{sec.results} we develop the application to the calcium isotope chain. In subsection \ref{sec.resultsA} we deal with the isospin independent model, while in subsection \ref{sec.resultsB} the isospin dependent approach is used to determine the neutron drip line. The last section \ref{sec.conclusions} is reserved for discussions and conclusions.

%----------------
\section{Formalism} \label{sec.method}

%_________________________________
\subsection{Exact pairing solution} \label{sec.methodA}
The Hamiltonian of a many-body system which includes pair scattering to the continuum may be written in terms of a set of negative and positive energy states, corresponding to bound and scattering states, respectively. For a constant pairing interaction the Hamiltonian is given by,
\begin{equation}
	H = \sum_\alpha \varepsilon_a c_\alpha^\dagger c_\alpha^{\;} - G \sum_{a m_\alpha} \sum_{c m_\zeta} c_\alpha^\dagger c_{\bar{\alpha}}^{\dagger} c_{\bar{\zeta}}^{\;} c_\zeta^{\;}
\end{equation}

\noindent  where $\varepsilon_a$ are the discrete energies with degeneracy \mbox{$\hat{j}^2_a=2j_a+1$}, $\alpha=\{a,m_\alpha\}=\{n_a, l_a, j_a,m_\alpha\}$ and \mbox{$c_{\bar{\alpha}}^{\dagger} = (-1)^{j_a-m_\alpha} c_{a,-m_\alpha}^\dagger$}. 

Following the derivation of Ref. \cite{2017IdBetan}, we may take the limit of the size of the spherical box to infinity, and keep only the physical relevant part of the single particle level density \cite{Beth1937II}. In this way, for a system with $N$ particles, we end up with $N_{\rm{pair}}=N/2$ couple equations, which take into account continuum correlations \cite{IdBetan2012,Pittel2015},
\begin{equation}
    \begin{split}
	1 &- \frac{G}{2} \sum_b^B \frac{\hat{j}^2_b}{2\varepsilon_b-E_{i}} 
	- \frac{1}{2} \int_0^\infty d\varepsilon \frac{Gg(\varepsilon)}{2\varepsilon - E_{i}}\\
	&- 2G \sum_{j\neq i}^{N_{\rm{pair}}} \frac{1}{E_{j} - E_{i}} = 0
	\end{split}
	\label{eq:Richeq}
\end{equation}
\noindent for $i=1,\cdots,N_{\rm{pair}}$, where $\varepsilon_b$ are the bound energy levels with quantum numbers $\{n_b, l_b, j_b\}$, $E_{i}$ are the Richardson energies which are parameters of the formalism, related to the many-body energy $E$ of the system \cite{VonDelft1999,Richardson1963} by,
\begin{equation}
	E(N_{\rm{pair}}) = \sum_{i=1}^{N_{\rm{pair}}} E_{i}
	\label{eq:eigenvalueequation}
\end{equation}
\noindent and $g(\varepsilon)$ is the Continuum Single Particle Level Density (CSPLD) \cite{Beth1937II},
\begin{equation} \label{eq.ge}
    g(\varepsilon) = \sum_c^{l_{\rm{max}}} \frac{\hat{j}^2_c}{\pi} \frac{d\delta_c}{d\varepsilon}
\end{equation}
\noindent where $l_{max}$ is an upper limit for the number of partial waves.

Notice, in Eq. (\ref{eq:Richeq}), that while the correlations between bound states are the same for all shells, the strength between continuum states is modulated by the CSPLD \cite{IdBetan2012P,IdBetan2012}.

The solution of the $N_{\rm{pair}}$ Richardson equations with the boundary conditions,
\begin{equation}
   \lim_{G\rightarrow0^+} E_{i} = 2 \varepsilon_{i}
\end{equation}
\noindent with $i=1,\cdots,N_{\rm{pair}}$ the lowest states, determine the ground-state energy of the pairing Hamiltonian of the $N=2N_{\rm{pair}}$ nucleus, where, the pair degeneracy $\hat{j}^2_i/2$ of the level $\varepsilon_{i}$ must be taken into account \cite{IdBetan2012}. For example, the isotope $^{44}$Ca, considered as a core $^{40}$Ca plus four neutrons, corresponds to solve two algebraic couple equations (\ref{eq:Richeq}) with the boundary conditions, 
$\lim_{G\rightarrow0^+} E_{1} = 2 \varepsilon_{1}$ and 
$\lim_{G\rightarrow0^+} E_{2} = 2 \varepsilon_{1}$, where $\varepsilon_{1}=\varepsilon_{f_{7/2}}$. In this case, the single particle energy limits are the same because the pair degeneracy of the shell $f_{7/2}$ is four. Then, the ground-state energy is given by Eq. (\ref{eq:eigenvalueequation}), i.e. $E=E_1+E_2$.

We will consider the independent and dependent isospin cases  \cite{1969Nilsson,IdBetan2017-R},
\begin{equation}\label{eq:pairingconstantiso}
	G = \frac{\chi_1}{A}(1-\chi_2I)
\end{equation}
where $I=\frac{N-Z}{A}$.

%_____________________________________________
\subsection{One and two-neutron separation energies} \label{sec.methodB}
The drip line becomes defined by the conditions $S_n \le 0$ and $S_{2n} \le 0$. Let us consider $A=A_{\rm{core}}+N$, where $A_{\rm{core}}$ is the inert core from where the mean-field Hamiltonian is set up, and $N=2N_{\rm{pair}}$. Then, the two-neutron separation energy from the Richardson formalism is given by,
\begin{equation} \label{eq.S2n}
  S_{2n}(A)=E(N_{\rm{pair}}-2) - E(N_{\rm{pair}})  
\end{equation}
with $E(N_{\rm{pair}})$ from Eq. (\ref{eq:eigenvalueequation}). While the one-neutron separation energy is calculated from the approximate equation \cite{Beiner1975,1996Dobaczewski},
\begin{equation} \label{eq.sn}
  S_n(A+1,Z) = - \lambda_F + \frac{\partial \lambda_F}{\partial N} - \Delta 
\end{equation}
with $\lambda_F$ and $\Delta$ the Fermi level and pairing gap, respectively, calculated in the blocking approximation, i.e.  $\lambda_F(2N_{\rm{pair}})$ and $\Delta(2N_{\rm{pair}})$; while $\frac{\partial \lambda_F}{\partial N}$ is calculated in \mbox{Sec. \ref{sec.lambda}}.

From the Richardson formalism, the Fermi level and the pairing gap can be calculated by combining the BCS equations with continuum spectrum \cite{IdBetan2012P},
\begin{align}
   	\Delta = \Delta_b + \Delta_c  
   	          &= \frac{G}{2}  \sum_b^B \hat{j}^2_b\;v_b \sqrt{1 - v_b^2} \,+ \nonumber \\
	                & + \frac{1}{2} \int_{0}^{\varepsilon_{\rm{max}}}
	                         d\varepsilon\;Gg(\varepsilon)\;
	                			v(\varepsilon) \sqrt{1 - v^2(\varepsilon)} \label{eq.delta}  \\
	 %%%
	N = N_b + N_c 
       &= \sum_b^B \frac{\hat{j}^2_b}{2} 
                   \left[ 1-\frac{\varepsilon_b - \lambda_F}
                           {\sqrt{(\varepsilon_b-\lambda_F)^2 + \Delta^2}} \right] \; + \nonumber \\
	             & + \int_0^{\varepsilon_{\rm{max}}}
	                   d\varepsilon\;\frac{g(\varepsilon)}{2}
	               \left[ 1-\frac{\varepsilon - \lambda_F}
	                      {\sqrt{(\varepsilon-\lambda_F)^2 + \Delta^2}} \right]  \label{eq.nbig}
\end{align}
and the occupation probabilities,
\begin{align}
   v_b^2 &= - G^2 \frac{d}{dG}
                 \left[\sum_{\nu=1}^{N_{\rm{pair}}}\frac{1}{2\varepsilon_b - E_\nu}\right] 
                 \label{eq.v2a} \\
   v^2(\varepsilon) &=  - G^2 \frac{d}{dG}
               \left[\sum_{\nu=1}^{N_{\rm{pair}}}\frac{1}{2\varepsilon - E_\nu}\right]    \label{eq.v2b}            
\end{align}
were we have extended the definition \cite{1964Richardson} to the continuum spectrum of energy, and we have introduced a cutoff $\varepsilon_{\rm{max}}$.

For a given nucleus $N=2N_{\rm{pair}}$, we solve the Richardson equations (\ref{eq:Richeq}) for many strengths $G$. Then, from \mbox{Eqs. (\ref{eq.v2a}) and (\ref{eq.v2b})} we calculated the occupation probabilities by finite differences. By substituting these results in Eq. (\ref{eq.delta}), we obtain the pairing gap. Finally, with this value of $\Delta$, we fit $\lambda_F$ from Eq. (\ref{eq.nbig}). In this way, the Fermi level and the pairing gap have been obtained for each even nucleus. Using these parameters in Eq. (\ref{eq.sn}) we get the one-neutron separation energy $S_n$ for the $A+1$ nucleus.

In the applications we also will show binding energy for the even $A=A_{\rm{core}}+2N_{\rm{pair}}$ and odd $A+1$ isotopes, given by,
\begin{align}
    E_{Bin}(A) &= E_{Bin}(A_{\rm{core}}) + E(N_{\rm{pair}}) \label{eq.eeven} \\
    E_{Bin}(A+1) &= E_{Bin}(A) + S_n(A+1) \label{eq.eodd}
\end{align}
were $ E_{Bin}(A_{\rm{core}})$ will be taken from experimental data.

%----------------
\section{Results } \label{sec.results}

%______________________________________
\subsection{Isospin independent model} \label{sec.resultsA}
We begin with the calculation of the drip line for the calcium isotopes in the isospin independent approximation.

\subsubsection{Single particle representation} 
Even when the solution of the reduced pairing Hamiltonian does not require the single-particle wave function of the mean-field but only the energies, in our formulation we make use of the single particle density Eq. (\ref{eq.ge}), which requires  the continuum eigenfunctions, and so, we need to define a mean-field. The Woods-Saxon and spin-orbit parameters were constrained by experimental data and $\chi^2$ optimization.

In this section we consider fixed strengths for the mean-field of the cores $^{40}$Ca and $^{48}$Ca. We will take the same reduced radius and diffuseness for both cores, in preparation for section \ref{sec.resultsB}, where the strengths of both cores will be joined smoothly. The reduced radius $r_0=1.28$ fm is extracted from the experimental neutron root-mean-square $r_n=3.555$ fm for $^{48}$Ca \cite{2018Zenihiro} and the relation $r_n=\sqrt{3/5}R$. For the diffuseness we take $a=0.75$ fm in order to get into consideration the enhancement of the nuclear size reported in Ref. \cite{2019Tanaka} which is justified by and increase in the surface diffuseness of the neutron density distribution. Finally, the strengths are optimized by $\chi^2$ using the Levenberg-Marquardt algorithm \cite{nr}. Due to the fragmentation of the single particle states in the nuclei $^{41}$Ca and $^{49}$Ca, we take as experimental energies, the average of the fragmented levels weighted with its respective spectroscopic factor \cite{2007Schwierz}. The optimized strength with their errors are shown in Table \ref{tab:parapote}.

\begin{table}[htb]
	\caption{\label{tab:parapote} Strength for the Woods-Saxon (MeV)  and spin-orbit (MeV fm) mean fields for the two model cores, with the error in parenthesis, and with $r_0=1.28$ fm and $a=0.75$ fm.}
  \begin{ruledtabular}
	\begin{tabular}{c | c | c}
		Core & Potential & $\mathcal{V}$\\
		\hline
		\multirow{2}{*}{$^{40}_{20}$Ca$_{20}^{\;}$}  & Woods-Saxon & $51.39(0.94)$\\
		& Spin-Orbit & $16.56(3.31)$\\
		\hline
		\multirow{2}{*}{$^{48}_{20}$Ca$_{28}^{\;}$}& Woods-Saxon & $45.39(1.47)$\\
		& Spin-Orbit & $18.36(6.03)$
	\end{tabular}
\end{ruledtabular}
\end{table}

\begin{figure*}[h!t]
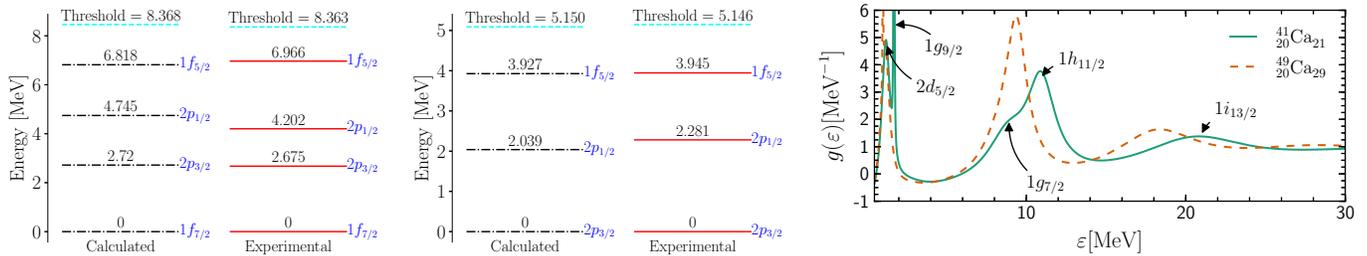

	\centering
	\begin{subfigure}{.3\textwidth}
		\resizebox{\textwidth}{!}{\input{Core40Ca.pgf}}
	\end{subfigure}%
	\begin{subfigure}{.3\textwidth}
		\resizebox{\textwidth}{!}{\input{Core48Ca.pgf}}
	\end{subfigure}%
	\begin{subfigure}{.42\textwidth}
		\resizebox{\textwidth}{!}{\input{LevelDensity41vs49_v2.pgf}}
	\end{subfigure}
	\vspace{-10mm}
	\caption{ \label{fig.basis} (Left and Center) Calculated and experimental \cite{nndcpag} levels of $^{41}$Ca and $^{49}$Ca, respectively. The dotted line shows the continuum threshold. (Right) Continuum single particle level density. The bumps are manifestation of the single particle resonances, labeled following the usual notation as for bound states.}
\end{figure*}

The left and center panels of \mbox{Fig. \ref{fig.basis}} compare the average experimental neutron levels of $^{41}$Ca and $^{49}$Ca \cite{nndcpag}, with that calculated using the code GAMOW \cite{Vertse1982} with the parameters of Table \ref{tab:parapote}. The right panel shows the continuum single-particle level density $g(\varepsilon)$ Eq. (\ref{eq.ge}), with $l_{\rm{max}}=15$. The scattering states were calculated using the code  ANTI \cite{Ixaru1995,1996Liotta}. The peaks are manifestation of the single particle resonances, which are labeled following the usual convention for bound-state shells. We observe that resonances move to the continuum threshold while they became narrower when changing from $^{41}$Ca to $^{49}$Ca. The figure shows a near degeneracy of the levels $1g_{9/2}$ and $2d_{5/2}$  for both nuclei \cite{2012Hamamoto}, which manifest as a single peak in the $^{49}$Ca. In section \ref{sec.resultsB} we will show the evolution of the single particle levels with $A$.

\subsubsection{Binding energy} 
Using the two single particle model spaces for the cores $^{40}$Ca and $^{48}$Ca, formed by the bound and continuum states of Fig. \ref{fig.basis}, we solve the Richardson equations (\ref{eq:Richeq}) for the calcium isotopes. Then, using Eqs. (\ref{eq:eigenvalueequation}) and (\ref{eq.eeven}) we calculate the binding energy of the even isotopes. 

The pairing strength $G$ is parametrized by Eq. (\ref{eq:pairingconstantiso}) with $\chi_2=0$. The reduced pairing strength $\chi_1$, for each core, were fixed in order to reproduce the experimental binding energy of two nuclei, one for the core $^{40}$Ca, and another for the core $^{48}$Ca. Table \ref{tab:chiadjust} shows the value of the parameter $\chi_1$ and compare the calculated and the experimental binding energy of the nuclei $^{50}$Ca and $^{54}$Ca used as reference. 
\begin{table}[ht]
\caption{ \label{tab:chiadjust} Parameters for isospin independent ($\chi_2=0$) pairing strength of Eq. (\ref{eq:pairingconstantiso}). The experimental binding energies are from Ref. \cite{Wang2017}.}
\begin{ruledtabular}
\begin{tabular}{c | c | c | c | c }
 Core & Nucleus & $E^{\rm{Exp}}_{Bin}$[MeV] & $E^{\rm{Cal}}_{Bin}$[MeV] & $\chi_1$[MeV]\\
		\hline
		&&&&\\[-0.8em]
 $^{40}$Ca  & $^{50}$Ca  & $-427.508(1)$ & $-427.508$ & $22.850$\\
		\hline
		&&&&\\[-0.8em]
 $^{48}$Ca  &	$^{54}$Ca & $-445.36(4)$ & $-445.367$ & $23.274$\\
\end{tabular}
\end{ruledtabular}
\end{table}

\begin{figure}[h!t] 
	\resizebox{.9\columnwidth}{!}{\input{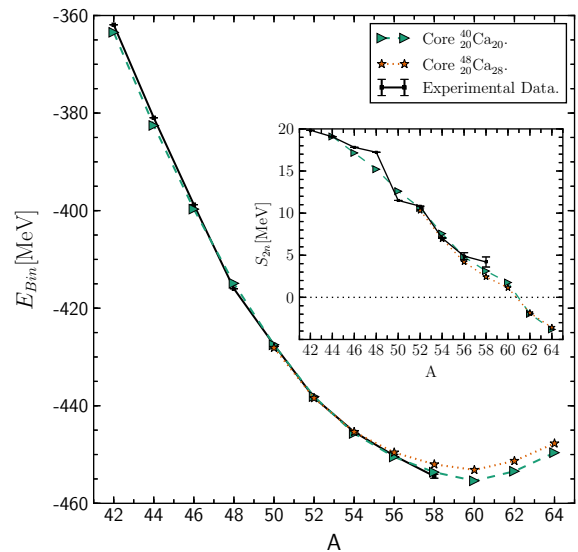}}
	\caption[T]{\label{fig:from40}
		Binding energies of the even calcium isotopes calculated using the isospin pairing strength of Table \ref{tab:chiadjust} for each one of the two model spaces. The experimental data were taken from \cite{Wang2017}. The inset shows the two-neutron separation energies.} 
\end{figure}

Using the reduced pairing strength $\chi_1$ of Table \ref{tab:chiadjust}, we calculate the binding energy of the calcium chain for each one of the model spaces, i.e. the one defined by the core $^{40}$Ca and the other by the core $^{48}$Ca. The results are shown in Figure \ref{fig:from40}. The two-neutron separation energy, calculated using Eq. (\ref{eq.S2n}), is shown in the inset. The results of both model spaces follow the experimental energy till the nucleus $^{54}$Ca, and then, the solutions using the core $^{40}$Ca does a better job. Both model spaces found the two-neutron drip line at the nucleus $^{60}$Ca, in concordance with \mbox{Refs. \cite{2012Hagen,2013Forssen,2018Tichai,2019Holt,2020Soma}.}

\begin{figure}[h!t] 
	\resizebox{.9\columnwidth}{!}{\input{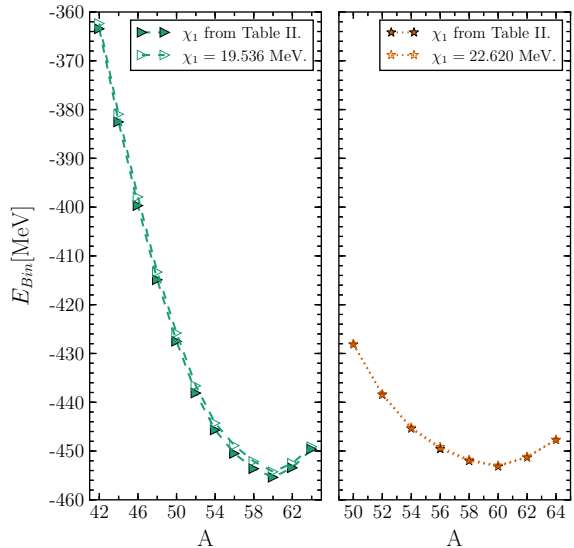}}
	\caption[T]{\label{fig:from40b}
		Binding energies of the even calcium isotopes for the model space with the core $^{40}$Ca (left) and the model space with the core $^{48}$Ca (right), for the two set of parametrization of the pairing strength as described in the text.} 
\end{figure}

Since the selection of the nuclei $^{50}$Ca and $^{54}$Ca (Table \ref{tab:chiadjust}) to fix the reduced pairing strength was arbitrary, we considered a second pair of reference nuclei, $^{44}$Ca and $^{52}$Ca, for the model spaces with core $^{40}$Ca and $^{48}$Ca, respectively. With the new pair of reduced pairing strengths $\chi_1$ we calculate the binding energy, and compare them with the previous one in  Fig. \ref{fig:from40b}. The new calculations found the two-neutron drip line at $^{60}$Ca, for both model spaces, in agreement with the previous parametrization.

Motivated by the analysis of Ref. \cite{2018Tarasov} and other theoretical predictions \cite{2019Neufcourt,2020Li}, we consider, in the next section, the dependence of isospin on the mean field and on the pairing force for the determination of the drip line.

%----------------
\subsection{Isospin dependent model} \label{sec.resultsB}
In this section we will consider the solution of the Richardson equations from the core $^{48}$Ca, with an isospin dependent single-particle model space and isospin dependent pairing strength.

%___________________________________________
\subsubsection{Single particle representation} \label{subsec.spr}
The single particle bound states and the CSPLD will change smoothly from isotope to isotope according to the following isospin dependent Woods-Saxon and spin-orbit strengths \cite{PhysRev.182.1190},
\begin{align}
	\mathcal{V}_{0} & = \eta_0 - \eta_1 I \label{eq:V0fit}\\
	\mathcal{V}_{SO} & = \eta_0^{SO} - \eta_1^{SO} I \label{eq:Vsofit}
\end{align}
with $I=\frac{N-Z}{A}$. The four parameters $\eta$, shown in Table \ref{table.mf2}, were fixed using the four strengths of Table \ref{tab:parapote} optimized by $\chi^2$ minimization in the previous section. 

\begin{table}[ht]
\caption{ \label{table.mf2} Parameters for the isospin dependent Woods-Saxon and spin-orbit strengths.}
\begin{ruledtabular}
\begin{tabular}{c | c | c }
   & Woods-Saxon [MeV] & spin-orbit [MeV fm] \\
		\hline
		&&\\[-0.8em]
 $\eta_0$  & $51.389$  & $16.564$  \\
		\hline
		&&\\[-0.8em]
 $\eta_1$  &	$36.005$ & $-10.759$ \\
\end{tabular}
\end{ruledtabular}
\end{table}

\begin{figure}[h!t] 
	\resizebox{.9\columnwidth}{!}{\input{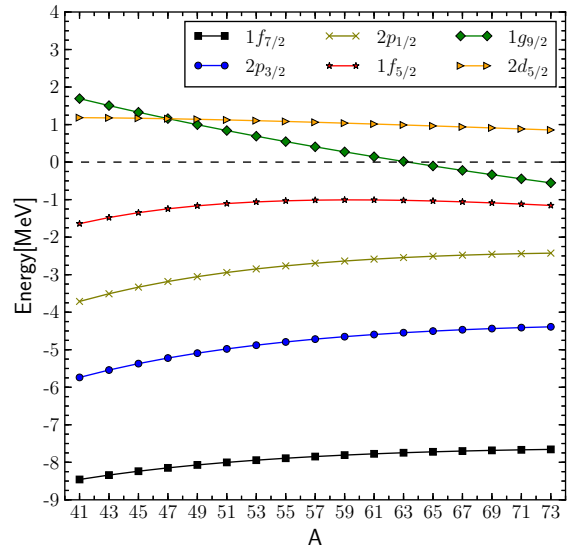}}
	\caption[T]{\label{fig:levels57}
		Evolution of the bound levels and the real part of the resonances $2d_{5/2}$ and $1g_{7/2}$, as a function of $A$ using $r_0=1.28$ fm, $a=0.75$ fm and the isospin dependent strength with the parameters of  Table \ref{table.mf2}.}
\end{figure}

The evolution of the bound levels of Fig. \ref{fig.basis} and the real part energy of the resonances $2d_{5/2}$ and $1g_{7/2}$, as a function of $A$ up to $^{73}$Ca, are shown in Fig. \ref{fig:levels57}. They were calculated using the code GAMOW \cite{Vertse1982}, with $r_0=1.28$ fm, $a=0.75$ fm and the isospin dependent strength Eqs. (\ref{eq:V0fit}) and (\ref{eq:Vsofit}), with the parameters of  Table \ref{table.mf2}. From the figure can be appreciated the inversion \cite{2012Hagen} and the near degeneracy \cite{2012Hamamoto} of the levels $2d_{5/2}$ and $1g_{7/2}$. The figure shows the transition of the state $g_{9/2}$ from a resonance to a bound state. This behavior seems to be a consequence of the increasing of the effective spin-orbit strength with $l$, and the enhancement of the centrifugal barrier, which is proportional to $l(l+1)$. These two factors are more pronounced for the $g_{9/2}$ shell. Figure \ref{fig:levels57} also shows two gaps between the shells $2p_{3/2}$-$2p_{1/2}$ and $2p_{1/2}$-$1f_{5/2}$, which are consistent with the shell closure of the nuclei $^{52}$Ca and $^{54}$Ca \cite{2018Michimasa,2018Leistenschneider,2019Liu,2020Leistenschneider,2020Li}. The weakening of shell closure at $^{60}$Ca, due to the tendency of the shell $g_{9/2}$, is in agreement with Ref. \cite{2020Li}, but we do not find a shell closure at $^{70}$Ca \cite{2020Li}. 
	
\begin{figure}[h!t]
	\resizebox{1.\columnwidth}{!}{\input{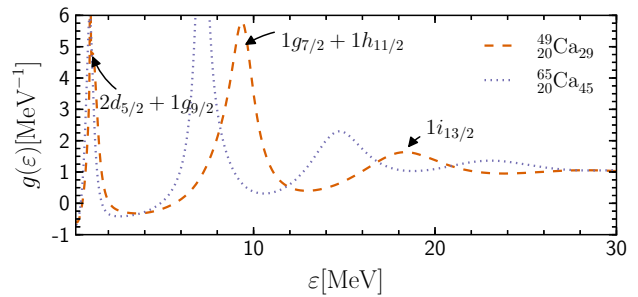}}
	\caption[T]{\label{fig:cspld55}
		Two examples of the continuum single particle level density used to solve the Richardson equations. The mean-field is the same as that used to construct Fig. \ref{fig:levels57}.}
\end{figure}

The continuum spectrum of energy enters the many-body calculation through the continuum single particle level density, which also smoothly changes from isotope to isotope. In Fig. \ref{fig:cspld55} we show, as an example, how the CSPLD profile changes from the nucleus $^{49}$Ca to the nucleus $^{65}$Ca. The resonant peaks move to the continuum threshold for increasing $A$.

%______________________________________________
\subsubsection{Two-neutron separation energy}
In this section we solve the Richardson equations (\ref{eq:Richeq}) for the even isotopes from $^{50}$Ca to $^{74}$Ca. The core is taken to be the nucleus $^{48}$Ca, with the model space as described in the previous section \ref{subsec.spr}. The isospin pairing strength $G$ is modeled by Eq. (\ref{eq:pairingconstantiso}), with the parameters $\chi_1$ and $\chi_2$ optimized to reproduce the experimental binding energy of the nuclei $^{54}$Ca and $^{58}$Ca, Table \ref{tab:chiadjustiso}.

\begin{table}[ht]
\caption{ \label{tab:chiadjustiso} Reduced isospin strengths $\chi_1$ and $\chi_2$ optimized with the experimental binding energy of the nuclei $^{54}$Ca and $^{58}$Ca.}
\begin{ruledtabular}
\centering
\begin{tabular}{ c  c  c  c  c }
   Nucl & $E^{\rm{Exp}}_{\rm{Bin}}$ (MeV) & $E^{\rm{Cal}}_{\rm{Bin}}$ (MeV) 
           & $\chi_1$ (MeV) & $\chi_2$ (MeV) \\
\hline
&&&&\\[-0.8em]
    $^{54}$Ca  & $-445.36(4)$ & $-445.363$  
       &\multirow{2}{*}{$16.314$}  
       &\multirow{2}{*}{$-1.108$}  \\
   $^{58}$Ca  & $-454.4(4)$ & $-454.400$ 
        &
        & 
\end{tabular}
\end{ruledtabular}
\end{table}
%%%%%%

\begin{figure}[h!t] 
	\resizebox{.9\columnwidth}{!}{\input{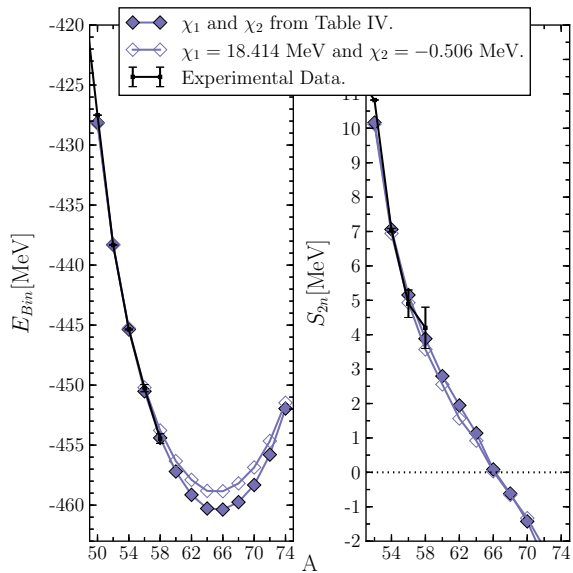}}
	\caption[T]{\label{fig:withcorrections} Binding energies and two-neutron separation energies from the model space and pairing strength isospin dependent. The result for two different sets of reduced pairing strengths $\chi_1$ and $\chi_2$ are given. The experimental data was taken from \cite{Wang2017}.}
\end{figure}

The calculated binding energy of the even isotopes and the two-neutron separation energy is shown in Fig. \ref{fig:withcorrections}. The figure shows that the last even isotope is $^{66}$Ca. This result is consistent with that of Ref. \cite{2019Neufcourt} which found the nuclei $^{66}$Ca and $^{68}$Ca to be bound, with a probability $67\%-84\%$. Reference \cite{2020Li} finds a pronounced smoothing of the binding energy for the isotopes $^{66}$Ca-$^{70}$Ca, with the drip line at the nucleus $^{70}$Ca.

Since the pair of nuclei used to fix the reduced pairing strengths $\chi_1$ and $\chi_2$ have nothing of particular, we repeated the calculation fixing the reduced pairing strengths using the experimental binding energy of the nuclei $^{52}$Ca and $^{56}$Ca. Figure \ref{fig:withcorrections} shows the calculation with the new pair of $\chi_1$ and $\chi_2$.  We observe a difference in the binding energy using the two different set of parameters, while there is a good agreement for two-neutron separation energy. The second parametrization also finds the drip line at the nucleus $^{66}$Ca.

%__________________________________________
\subsubsection{Pairing in the continuum} \label{sec.lambda}
\begin{figure}[!ht]
	\resizebox{.9\columnwidth}{!}{\input{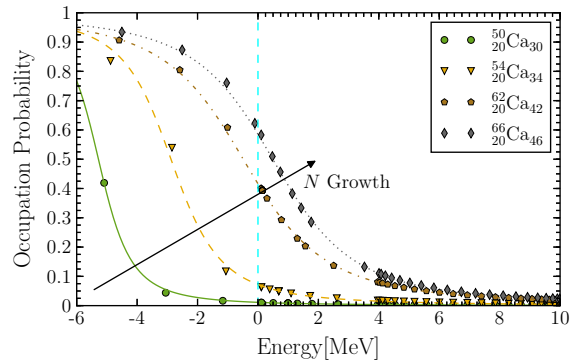}}
	\caption[T]{\label{fig:proboccup}
		Occupation Probability for some selected even calcium isotopes. The lines correspond to fitted curves using the BCS \cite{1957Bcs} distribution. The vertical dashed line indicates the continuum threshold.}
\end{figure}

By solving the Richardson equations for the pairing strength with the parameters of Table \ref{tab:chiadjustiso}, we calculate the occupation probability $v^2_b$ and $v^2(\varepsilon)$ for the bound and continuum states from \mbox{Eqs. (\ref{eq.v2a}) and (\ref{eq.v2b}),} respectively. Figure \ref{fig:proboccup} shows some examples; it can be observed how the occupation probabilities of the continuum levels, $\varepsilon > 0$, monotonically increase as the number of particles increases.

\begin{figure}[!ht] 
	\resizebox{1.\columnwidth}{!}{\input{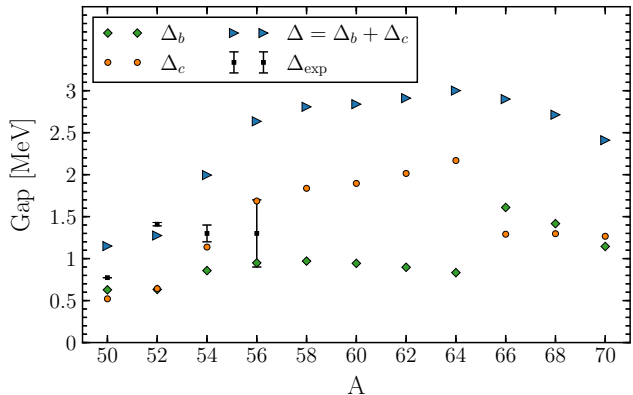}} 
	\caption[T]{\label{fig:gap}
		Pairing gap (\ref{eq.delta}) discriminated by bound $\Delta_b$ and continuum $\Delta_c$ contributions. The experimental gap was calculated using the three-point formula \cite{Changizi2015}, with binding energies from Ref. \cite{Wang2017}.
	}
\end{figure}

With the calculated occupation probabilities we get the discrete $\Delta_b$ and continuum $\Delta_c$ gap parameters from Eq. (\ref{eq.delta}) with $\varepsilon_{\rm{max}}=100$ MeV. Figure \ref{fig:gap} shows the total gap $\Delta$ discriminate by the discrete and continuum parts. The profile of the total gap is the usual for a strong pairing. Using the three-point \mbox{formula \cite{Changizi2015},} with the experimental binding energies from Ref. \cite{Wang2017}, we calculate the experimental gap $\Delta_{\rm{exp}}$; except for the nucleus $^{52}$Ca, our gap are greater than the experimental one. The figure shows that $\Delta_c$ increases while $\Delta_b$ remains more or less constant up to the isotope $^{64}$Ca, where both suddenly change, but, the total pairing gap $\Delta$ remains smooth. The abrupt change of $\Delta_b$ and $\Delta_c$ is due that the state $1g_{9/2}$ becomes a bound state, as it can be seen in Fig. \ref{fig:levels57}.

\begin{figure}[h!t] 
	\resizebox{.8\columnwidth}{!}{\input{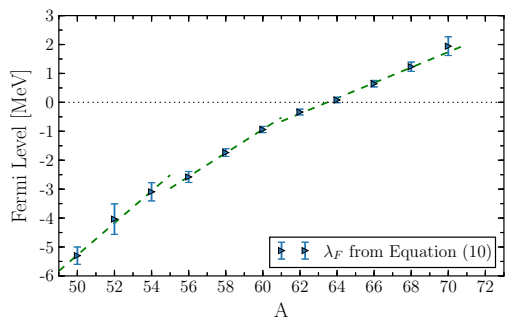}} 
	\caption[T]{\label{fig:lambdaF}
		Fermi level for even calcium isotopes calculated as described in the text. The lines were obtained by linear regression.}
\end{figure}

Finally, with the calculated gap $\Delta$, we determine the corresponding Fermi level by optimizing the parameter $\lambda_F$ in Eq. (\ref{eq.nbig}) using Levenberg-Marquardt algorithm \cite{root}. Figure \ref{fig:lambdaF} shows the optimized values with their errors.

For the determination of the one-neutron separation energy Eq. (\ref{eq.sn}), we also need $\partial \lambda_F/ \partial N$. Using linear regression we get (in unit of MeV), 
\begin{equation}\label{eq.sn2}
  \frac{\partial\lambda_F}{\partial N} =
  \begin{cases} 
        0.55(11) & 50 \le A \le 54 \\
        0.41(5)   & 56 \le A \le 60 \\ 
        0.27(3)  & 62 \ge A        
  \end{cases}
\end{equation}
Since the Fermi level measures the change of energy with $N$, it shows that this magnitude is smaller approaching to the drip line.

%__________________________________________
\subsubsection{One-neutron separation energy}
\begin{figure}[!ht] 
	\resizebox{.9\columnwidth}{!}{\input{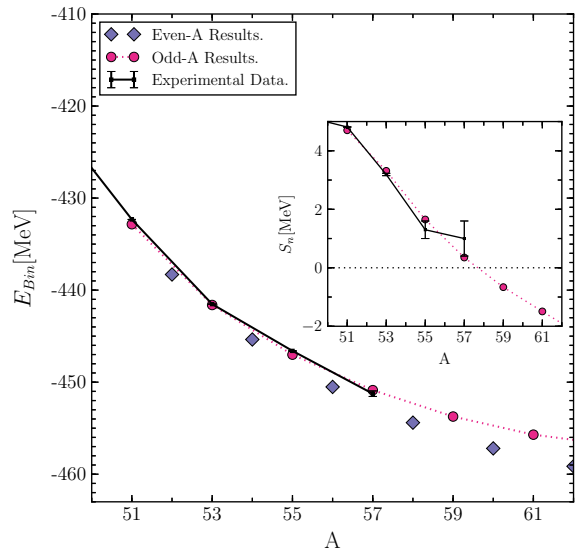}}
	\caption[T]{\label{fig:oddA}
		Binding energy of the even and odd calcium isotopes. The experimental data was taken from \cite{Wang2017}.}
\end{figure}

To complete the determination of the drip line we will calculate the one-neutron separation energy from \mbox{Eqs. (\ref{eq.sn}) and (\ref{eq.sn2})} and the magnitudes of the previous sub-section. Then, using Eq. (\ref{eq.eodd}) we evaluate the binding energy for the odd calcium isotopes, which is shown in Fig. \ref{fig:oddA}. The usual staggering, mounting onto the parabola-like curve, can be observed. The inset shows the one-neutron separation energy, the comparison with the experimental data shows a good agreement. We found that the one-neutron drip line happens to be at $^{57}$Ca, in agreement with ab initio models \cite{2012Hagen,2013Forssen}, and the Gamow Shell Model \cite{2020Li}, but in disagreement with the experimental result of Ref. \cite{2018Tarasov} which found that the nucleus $^{59}$Ca is also bound.

%----------------
\section{Discussion and Conclusions} \label{sec.conclusions}
We have calculated the one- and two-neutron separation energy of the calcium isotopes from the exact solution of the pairing Hamiltonian. While the two-neutron separation energy is obtained straightforward from the Richardson solution, the one-neutron separation energy was calculated using the pairing gap $\Delta$ and Fermi level $\lambda_F$, borrowed from the BCS formalism. The occupation probabilities needed to calculate $\Delta$ and $\lambda_F$ were obtained from the exact solution of the pairing Hamiltonian by finite difference. The correlations with the continuum spectrum of energy was taken into account through the single-particle density. Outcomes from isospin independent and dependent mean-field and pairing were investigated.

The evolution of the single particle levels shows an inversion of the shells $2d_{5/2}$ and $1g_{9/2}$ at the beginning of the chain, as reported in Ref. \cite{2012Hagen}; and then, a near degeneracy, as the one reported in \cite{2012Hamamoto} for deformed nuclei. Finally, the original order is reversed to the usual shell ordering, with the shell $1g_{9/2}$ becoming a bound state, at the time that the shell $2d_{5/2}$ remains in the continuum. The displacement of the single particle levels shows a shell closure for the calcium isotopes with $N=32$ and $N=34$. The intrusion of the shell $1g_{9/2}$ from the continuum slightly hinders a closure for $N=40$, and prevents a closure at $N=50$. The influence of deformations upon the level structures in very rich neutron nuclei is expected to be important, in particular, more experimental structure information on the calcium isotopes is expected in the near future.

Our calculation found the nucleus $^{57}$Ca as the last bound odd isotope, in agreement with \cite{2020Li}, but in disagreement with the experimental finding reported in Ref. \cite{2018Tarasov}, probably due and overestimation of the pairing gap.

The results from the isospin independent formulation shows that  $^{60}$Ca is the last bound calcium isotope. Similar result was found using Bogoliubov perturbation formalism \cite{2018Tichai} and self-consistent Green's function \cite{2020Soma}. By including the isospin dependence in the mean-field and pairing strength, drip line is extended to $^{66}$Ca. This result is smaller than the predictions from the Bayesian Model Averaging \cite{2019Neufcourt} and the Gamow Shell Model \cite{2020Li}, which allocate the drip line around $^{68}$Ca-$^{70}$Ca. %Our calculation found the nucleus $^{57}$Ca as the last bound odd isotope, in agreement with \cite{2020Li}, but in disagreement with the experimental finding reported in Ref. \cite{2018Tarasov}, probably due and overestimation of the pairing gap.

%The outcome of this paper shows the importance of the isospin dependence on the pairing strength, besides the usual $1/A$ dependence, i.e., $(N-Z)/A^2$. It also shows the aptitude of the exact pairing solution to describe the one and two-neutron drip lines in the calcium isotope chain.
The outcome of this paper shows the ability of the exact pairing formalism to describe one and two neutron drip lines in the calcium isotope chain. In the current state of knowledge, all three models reproduce the known data equally well. This places an uncertainty in our prediction for the two neutron drip line, with $^{60}$Ca or $^{66}$Ca depending on whether the independent or dependent isospin model is considered.% New experimental data will resolved it.

%-----------------------
\begin{acknowledgments}
This work has been supported by the National Council of Research PIP-625 and the University of Rosario ING588, Argentina.
%Authors are grateful to Prof. ... for valuable discussions.
\end{acknowledgments}

%-----------------------
%\bibliographystyle{unsrturl85}  
%\bibliography{BibCalcium}

\begin{thebibliography}{64}%
\makeatletter
\providecommand \@ifxundefined [1]{%
 \@ifx{#1\undefined}
}%
\providecommand \@ifnum [1]{%
 \ifnum #1\expandafter \@firstoftwo
 \else \expandafter \@secondoftwo
 \fi
}%
\providecommand \@ifx [1]{%
 \ifx #1\expandafter \@firstoftwo
 \else \expandafter \@secondoftwo
 \fi
}%
\providecommand \natexlab [1]{#1}%
\providecommand \enquote  [1]{``#1''}%
\providecommand \bibnamefont  [1]{#1}%
\providecommand \bibfnamefont [1]{#1}%
\providecommand \citenamefont [1]{#1}%
\providecommand \href@noop [0]{\@secondoftwo}%
\providecommand \href [0]{\begingroup \@sanitize@url \@href}%
\providecommand \@href[1]{\@@startlink{#1}\@@href}%
\providecommand \@@href[1]{\endgroup#1\@@endlink}%
\providecommand \@sanitize@url [0]{\catcode `\\12\catcode `\$12\catcode
  `\&12\catcode `\#12\catcode `\^12\catcode `\_12\catcode `\%12\relax}%
\providecommand \@@startlink[1]{}%
\providecommand \@@endlink[0]{}%
\providecommand \url  [0]{\begingroup\@sanitize@url \@url }%
\providecommand \@url [1]{\endgroup\@href {#1}{\urlprefix }}%
\providecommand \urlprefix  [0]{URL }%
\providecommand \Eprint [0]{\href }%
\providecommand \doibase [0]{http://dx.doi.org/}%
\providecommand \selectlanguage [0]{\@gobble}%
\providecommand \bibinfo  [0]{\@secondoftwo}%
\providecommand \bibfield  [0]{\@secondoftwo}%
\providecommand \translation [1]{[#1]}%
\providecommand \BibitemOpen [0]{}%
\providecommand \bibitemStop [0]{}%
\providecommand \bibitemNoStop [0]{.\EOS\space}%
\providecommand \EOS [0]{\spacefactor3000\relax}%
\providecommand \BibitemShut  [1]{\csname bibitem#1\endcsname}%
\let\auto@bib@innerbib\@empty
%</preamble>
\bibitem [{\citenamefont {Ahn}\ \emph {et~al.}(2019)\citenamefont {Ahn},
  \citenamefont {Fukuda}, \citenamefont {Geissel}, \citenamefont {Inabe},
  \citenamefont {Iwasa}, \citenamefont {Kubo}, \citenamefont {Kusaka},
  \citenamefont {Morrissey}, \citenamefont {Murai}, \citenamefont {Nakamura},
  \citenamefont {Ohtake}, \citenamefont {Otsu}, \citenamefont {Sato},
  \citenamefont {Sherrill}, \citenamefont {Shimizu} \emph {et~al.}}]{2019Ahn}%
  \BibitemOpen
  \bibfield  {author} {\bibinfo {author} {\bibfnamefont {D.~S.}\ \bibnamefont
  {Ahn}}, \bibinfo {author} {\bibfnamefont {N.}~\bibnamefont {Fukuda}},
  \bibinfo {author} {\bibfnamefont {H.}~\bibnamefont {Geissel}}, \bibinfo
  {author} {\bibfnamefont {N.}~\bibnamefont {Inabe}}, \bibinfo {author}
  {\bibfnamefont {N.}~\bibnamefont {Iwasa}}, \bibinfo {author} {\bibfnamefont
  {T.}~\bibnamefont {Kubo}}, \bibinfo {author} {\bibfnamefont {K.}~\bibnamefont
  {Kusaka}}, \bibinfo {author} {\bibfnamefont {D.~J.}\ \bibnamefont
  {Morrissey}}, \bibinfo {author} {\bibfnamefont {D.}~\bibnamefont {Murai}},
  \bibinfo {author} {\bibfnamefont {T.}~\bibnamefont {Nakamura}}, \bibinfo
  {author} {\bibfnamefont {M.}~\bibnamefont {Ohtake}}, \bibinfo {author}
  {\bibfnamefont {H.}~\bibnamefont {Otsu}}, \bibinfo {author} {\bibfnamefont
  {H.}~\bibnamefont {Sato}}, \bibinfo {author} {\bibfnamefont {B.~M.}\
  \bibnamefont {Sherrill}}, \bibinfo {author} {\bibfnamefont {Y.}~\bibnamefont
  {Shimizu}},  \emph {et~al.},\ }\href {\doibase
  10.1103/PhysRevLett.123.212501} {\bibfield  {journal} {\bibinfo  {journal}
  {Physical Review Letters}\ }\textbf {\bibinfo {volume} {123}},\ \bibinfo
  {pages} {212501} (\bibinfo {year} {2019})}\BibitemShut {NoStop}%
\bibitem [{\citenamefont {Artukh}\ \emph {et~al.}(1970)\citenamefont {Artukh},
  \citenamefont {Avdeichikov}, \citenamefont {Chelnokov}, \citenamefont
  {Gridnev}, \citenamefont {Mikheev}, \citenamefont {Vakatov}, \citenamefont
  {Volkov},\ and\ \citenamefont {Wilczynski}}]{1970Artukh}%
  \BibitemOpen
  \bibfield  {author} {\bibinfo {author} {\bibfnamefont {A.~G.}\ \bibnamefont
  {Artukh}}, \bibinfo {author} {\bibfnamefont {V.~V.}\ \bibnamefont
  {Avdeichikov}}, \bibinfo {author} {\bibfnamefont {L.~P.}\ \bibnamefont
  {Chelnokov}}, \bibinfo {author} {\bibfnamefont {G.~F.}\ \bibnamefont
  {Gridnev}}, \bibinfo {author} {\bibfnamefont {V.~L.}\ \bibnamefont
  {Mikheev}}, \bibinfo {author} {\bibfnamefont {V.~I.}\ \bibnamefont
  {Vakatov}}, \bibinfo {author} {\bibfnamefont {V.~V.}\ \bibnamefont {Volkov}},
  \ and\ \bibinfo {author} {\bibfnamefont {J.}~\bibnamefont {Wilczynski}},\
  }\href {\doibase 10.1016/0370-2693(70)90332-1} {\bibfield  {journal}
  {\bibinfo  {journal} {Physics Letters B}\ }\textbf {\bibinfo {volume} {32}},\
  \bibinfo {pages} {43} (\bibinfo {year} {1970})}\BibitemShut {NoStop}%
\bibitem [{\citenamefont {Langevin}\ \emph {et~al.}(1985)\citenamefont
  {Langevin}, \citenamefont {Quiniou}, \citenamefont {Bernas}, \citenamefont
  {Galin}, \citenamefont {Jacmart}, \citenamefont {Naulin}, \citenamefont
  {Pougheon}, \citenamefont {Anne}, \citenamefont {D\'etraz}, \citenamefont
  {Guerreau}, \citenamefont {Guillemaud-Mueller},\ and\ \citenamefont
  {Mueller}}]{1985Langevin}%
  \BibitemOpen
  \bibfield  {author} {\bibinfo {author} {\bibfnamefont {M.}~\bibnamefont
  {Langevin}}, \bibinfo {author} {\bibfnamefont {E.}~\bibnamefont {Quiniou}},
  \bibinfo {author} {\bibfnamefont {M.}~\bibnamefont {Bernas}}, \bibinfo
  {author} {\bibfnamefont {J.}~\bibnamefont {Galin}}, \bibinfo {author}
  {\bibfnamefont {J.}~\bibnamefont {Jacmart}}, \bibinfo {author} {\bibfnamefont
  {F.}~\bibnamefont {Naulin}}, \bibinfo {author} {\bibfnamefont
  {F.}~\bibnamefont {Pougheon}}, \bibinfo {author} {\bibfnamefont
  {R.}~\bibnamefont {Anne}}, \bibinfo {author} {\bibfnamefont {C.}~\bibnamefont
  {D\'etraz}}, \bibinfo {author} {\bibfnamefont {D.}~\bibnamefont {Guerreau}},
  \bibinfo {author} {\bibfnamefont {D.}~\bibnamefont {Guillemaud-Mueller}}, \
  and\ \bibinfo {author} {\bibfnamefont {A.}~\bibnamefont {Mueller}},\ }\href
  {\doibase 10.1016/0370-2693(85)90140-6} {\bibfield  {journal} {\bibinfo
  {journal} {Physics Letters B}\ }\textbf {\bibinfo {volume} {150}},\ \bibinfo
  {pages} {71} (\bibinfo {year} {1985})}\BibitemShut {NoStop}%
\bibitem [{\citenamefont {Guillemaud-Mueller}\ \emph
  {et~al.}(1990)\citenamefont {Guillemaud-Mueller}, \citenamefont {Jacmart},
  \citenamefont {Kashy}, \citenamefont {Latimier}, \citenamefont {Mueller},
  \citenamefont {Pougheon}, \citenamefont {Richard}, \citenamefont
  {Penionzhkevich}, \citenamefont {Artuhk}, \citenamefont {Belozyorov},
  \citenamefont {Lukyanov}, \citenamefont {Anne}, \citenamefont {Bricault},
  \citenamefont {D\'etraz}, \citenamefont {Lewitowicz} \emph
  {et~al.}}]{1990Mueller}%
  \BibitemOpen
  \bibfield  {author} {\bibinfo {author} {\bibfnamefont {D.}~\bibnamefont
  {Guillemaud-Mueller}}, \bibinfo {author} {\bibfnamefont {J.~C.}\ \bibnamefont
  {Jacmart}}, \bibinfo {author} {\bibfnamefont {E.}~\bibnamefont {Kashy}},
  \bibinfo {author} {\bibfnamefont {A.}~\bibnamefont {Latimier}}, \bibinfo
  {author} {\bibfnamefont {A.~C.}\ \bibnamefont {Mueller}}, \bibinfo {author}
  {\bibfnamefont {F.}~\bibnamefont {Pougheon}}, \bibinfo {author}
  {\bibfnamefont {A.}~\bibnamefont {Richard}}, \bibinfo {author} {\bibfnamefont
  {Y.~E.}\ \bibnamefont {Penionzhkevich}}, \bibinfo {author} {\bibfnamefont
  {A.~G.}\ \bibnamefont {Artuhk}}, \bibinfo {author} {\bibfnamefont {A.~V.}\
  \bibnamefont {Belozyorov}}, \bibinfo {author} {\bibfnamefont {S.~M.}\
  \bibnamefont {Lukyanov}}, \bibinfo {author} {\bibfnamefont {R.}~\bibnamefont
  {Anne}}, \bibinfo {author} {\bibfnamefont {P.}~\bibnamefont {Bricault}},
  \bibinfo {author} {\bibfnamefont {C.}~\bibnamefont {D\'etraz}}, \bibinfo
  {author} {\bibfnamefont {M.}~\bibnamefont {Lewitowicz}},  \emph {et~al.},\
  }\href {\doibase 10.1103/PhysRevC.41.937} {\bibfield  {journal} {\bibinfo
  {journal} {Physical Review C}\ }\textbf {\bibinfo {volume} {41}},\ \bibinfo
  {pages} {937} (\bibinfo {year} {1990})}\BibitemShut {NoStop}%
\bibitem [{\citenamefont {Lunderberg}\ \emph {et~al.}(2012)\citenamefont
  {Lunderberg}, \citenamefont {DeYoung}, \citenamefont {Kohley}, \citenamefont
  {Attanayake}, \citenamefont {Baumann}, \citenamefont {Bazin}, \citenamefont
  {Christian}, \citenamefont {Divaratne}, \citenamefont {Grimes}, \citenamefont
  {Haagsma}, \citenamefont {Finck}, \citenamefont {Frank}, \citenamefont
  {Luther}, \citenamefont {Mosby}, \citenamefont {Nagi} \emph
  {et~al.}}]{2012Lunderberg}%
  \BibitemOpen
  \bibfield  {author} {\bibinfo {author} {\bibfnamefont {E.}~\bibnamefont
  {Lunderberg}}, \bibinfo {author} {\bibfnamefont {P.~A.}\ \bibnamefont
  {DeYoung}}, \bibinfo {author} {\bibfnamefont {Z.}~\bibnamefont {Kohley}},
  \bibinfo {author} {\bibfnamefont {H.}~\bibnamefont {Attanayake}}, \bibinfo
  {author} {\bibfnamefont {T.}~\bibnamefont {Baumann}}, \bibinfo {author}
  {\bibfnamefont {D.}~\bibnamefont {Bazin}}, \bibinfo {author} {\bibfnamefont
  {G.}~\bibnamefont {Christian}}, \bibinfo {author} {\bibfnamefont
  {D.}~\bibnamefont {Divaratne}}, \bibinfo {author} {\bibfnamefont {S.~M.}\
  \bibnamefont {Grimes}}, \bibinfo {author} {\bibfnamefont {A.}~\bibnamefont
  {Haagsma}}, \bibinfo {author} {\bibfnamefont {J.~E.}\ \bibnamefont {Finck}},
  \bibinfo {author} {\bibfnamefont {N.}~\bibnamefont {Frank}}, \bibinfo
  {author} {\bibfnamefont {B.}~\bibnamefont {Luther}}, \bibinfo {author}
  {\bibfnamefont {S.}~\bibnamefont {Mosby}}, \bibinfo {author} {\bibfnamefont
  {T.}~\bibnamefont {Nagi}},  \emph {et~al.},\ }\href {\doibase
  10.1103/PhysRevLett.108.142503} {\bibfield  {journal} {\bibinfo  {journal}
  {Physical Review Letters}\ }\textbf {\bibinfo {volume} {108}},\ \bibinfo
  {pages} {142503} (\bibinfo {year} {2012})}\BibitemShut {NoStop}%
\bibitem [{\citenamefont {Otsuka}\ \emph {et~al.}(2010)\citenamefont {Otsuka},
  \citenamefont {Suzuki}, \citenamefont {Holt}, \citenamefont {Schwenk},\ and\
  \citenamefont {Akaishi}}]{2010Otsuka}%
  \BibitemOpen
  \bibfield  {author} {\bibinfo {author} {\bibfnamefont {T.}~\bibnamefont
  {Otsuka}}, \bibinfo {author} {\bibfnamefont {T.}~\bibnamefont {Suzuki}},
  \bibinfo {author} {\bibfnamefont {J.~D.}\ \bibnamefont {Holt}}, \bibinfo
  {author} {\bibfnamefont {A.}~\bibnamefont {Schwenk}}, \ and\ \bibinfo
  {author} {\bibfnamefont {Y.}~\bibnamefont {Akaishi}},\ }\href {\doibase
  10.1103/PhysRevLett.105.032501} {\bibfield  {journal} {\bibinfo  {journal}
  {Physical Review Letters}\ }\textbf {\bibinfo {volume} {105}},\ \bibinfo
  {pages} {032501} (\bibinfo {year} {2010})}\BibitemShut {NoStop}%
\bibitem [{\citenamefont {Brink}\ and\ \citenamefont
  {Broglia}(2007)}]{DavidM.Brink938}%
  \BibitemOpen
  \bibfield  {author} {\bibinfo {author} {\bibfnamefont {D.~M.}\ \bibnamefont
  {Brink}}\ and\ \bibinfo {author} {\bibfnamefont {R.~A.}\ \bibnamefont
  {Broglia}},\ }\href {\doibase 10.1017/CBO9780511534911} {\emph {\bibinfo
  {title} {Nuclear Superfluidity: Pairing in Finite Systems}}}\ (\bibinfo
  {publisher} {Cambridge University Press},\ \bibinfo {year}
  {2007})\BibitemShut {NoStop}%
\bibitem [{\citenamefont {Forss\'en}\ \emph {et~al.}(2013)\citenamefont
  {Forss\'en}, \citenamefont {Hagen}, \citenamefont {Hjorth-Jensen},
  \citenamefont {Nazarewicz},\ and\ \citenamefont {Rotureau}}]{2013Forssen}%
  \BibitemOpen
  \bibfield  {author} {\bibinfo {author} {\bibfnamefont {C.}~\bibnamefont
  {Forss\'en}}, \bibinfo {author} {\bibfnamefont {G.}~\bibnamefont {Hagen}},
  \bibinfo {author} {\bibfnamefont {M.}~\bibnamefont {Hjorth-Jensen}}, \bibinfo
  {author} {\bibfnamefont {W.}~\bibnamefont {Nazarewicz}}, \ and\ \bibinfo
  {author} {\bibfnamefont {J.}~\bibnamefont {Rotureau}},\ }\href {\doibase
  10.1088/0031-8949/2013/T152/014022} {\bibfield  {journal} {\bibinfo
  {journal} {Physica Scripta}\ }\textbf {\bibinfo {volume} {T152}},\ \bibinfo
  {pages} {014022} (\bibinfo {year} {2013})}\BibitemShut {NoStop}%
\bibitem [{\citenamefont {Erler}\ \emph {et~al.}(2012)\citenamefont {Erler},
  \citenamefont {Birge}, \citenamefont {Kortelainen}, \citenamefont
  {Nazarewicz}, \citenamefont {Olsen}, \citenamefont {Perhac},\ and\
  \citenamefont {Stoitsov}}]{2012Erler}%
  \BibitemOpen
  \bibfield  {author} {\bibinfo {author} {\bibfnamefont {J.}~\bibnamefont
  {Erler}}, \bibinfo {author} {\bibfnamefont {N.}~\bibnamefont {Birge}},
  \bibinfo {author} {\bibfnamefont {M.}~\bibnamefont {Kortelainen}}, \bibinfo
  {author} {\bibfnamefont {W.}~\bibnamefont {Nazarewicz}}, \bibinfo {author}
  {\bibfnamefont {E.}~\bibnamefont {Olsen}}, \bibinfo {author} {\bibfnamefont
  {A.}~\bibnamefont {Perhac}}, \ and\ \bibinfo {author} {\bibfnamefont
  {M.}~\bibnamefont {Stoitsov}},\ }\href {\doibase 10.1038/nature11188}
  {\bibfield  {journal} {\bibinfo  {journal} {Nature}\ }\textbf {\bibinfo
  {volume} {486}},\ \bibinfo {pages} {509} (\bibinfo {year}
  {2012})}\BibitemShut {NoStop}%
\bibitem [{\citenamefont {Johnson}\ \emph {et~al.}(2019)\citenamefont
  {Johnson}, \citenamefont {Launey}, \citenamefont {Auerbach}, \citenamefont
  {Bacca}, \citenamefont {Barrett}, \citenamefont {Brune}, \citenamefont
  {Caprio}, \citenamefont {Descouvemont}, \citenamefont {Dickhoff},
  \citenamefont {Elster}, \citenamefont {Fasano}, \citenamefont {Fossez},
  \citenamefont {Hergert}, \citenamefont {Hjorth-Jensenm}, \citenamefont
  {Hlophe} \emph {et~al.}}]{2020Johnson}%
  \BibitemOpen
  \bibfield  {author} {\bibinfo {author} {\bibfnamefont {C.~W.}\ \bibnamefont
  {Johnson}}, \bibinfo {author} {\bibfnamefont {K.~D.}\ \bibnamefont {Launey}},
  \bibinfo {author} {\bibfnamefont {N.}~\bibnamefont {Auerbach}}, \bibinfo
  {author} {\bibfnamefont {S.}~\bibnamefont {Bacca}}, \bibinfo {author}
  {\bibfnamefont {B.~R.}\ \bibnamefont {Barrett}}, \bibinfo {author}
  {\bibfnamefont {C.}~\bibnamefont {Brune}}, \bibinfo {author} {\bibfnamefont
  {M.~A.}\ \bibnamefont {Caprio}}, \bibinfo {author} {\bibfnamefont
  {P.}~\bibnamefont {Descouvemont}}, \bibinfo {author} {\bibfnamefont {W.~H.}\
  \bibnamefont {Dickhoff}}, \bibinfo {author} {\bibfnamefont {C.}~\bibnamefont
  {Elster}}, \bibinfo {author} {\bibfnamefont {P.~J.}\ \bibnamefont {Fasano}},
  \bibinfo {author} {\bibfnamefont {K.}~\bibnamefont {Fossez}}, \bibinfo
  {author} {\bibfnamefont {H.}~\bibnamefont {Hergert}}, \bibinfo {author}
  {\bibfnamefont {M.}~\bibnamefont {Hjorth-Jensenm}}, \bibinfo {author}
  {\bibfnamefont {L.}~\bibnamefont {Hlophe}},  \emph {et~al.},\ }\href@noop {}
  {\  (\bibinfo {year} {2019})},\ \Eprint {http://arxiv.org/abs/1912.00451}
  {arXiv:1912.00451 [nucl-th]} \BibitemShut {NoStop}%
\bibitem [{\citenamefont {Tarasov}\ \emph {et~al.}(2018)\citenamefont
  {Tarasov}, \citenamefont {Ahn}, \citenamefont {Bazin}, \citenamefont
  {Fukuda}, \citenamefont {Gade}, \citenamefont {Hausmann}, \citenamefont
  {Inabe}, \citenamefont {Ishikawa}, \citenamefont {Iwasa}, \citenamefont
  {Kawata}, \citenamefont {Komatsubara}, \citenamefont {Kubo}, \citenamefont
  {Kusaka}, \citenamefont {Morrissey}, \citenamefont {Ohtake} \emph
  {et~al.}}]{2018Tarasov}%
  \BibitemOpen
  \bibfield  {author} {\bibinfo {author} {\bibfnamefont {O.~B.}\ \bibnamefont
  {Tarasov}}, \bibinfo {author} {\bibfnamefont {D.~S.}\ \bibnamefont {Ahn}},
  \bibinfo {author} {\bibfnamefont {D.}~\bibnamefont {Bazin}}, \bibinfo
  {author} {\bibfnamefont {N.}~\bibnamefont {Fukuda}}, \bibinfo {author}
  {\bibfnamefont {A.}~\bibnamefont {Gade}}, \bibinfo {author} {\bibfnamefont
  {M.}~\bibnamefont {Hausmann}}, \bibinfo {author} {\bibfnamefont
  {N.}~\bibnamefont {Inabe}}, \bibinfo {author} {\bibfnamefont
  {S.}~\bibnamefont {Ishikawa}}, \bibinfo {author} {\bibfnamefont
  {N.}~\bibnamefont {Iwasa}}, \bibinfo {author} {\bibfnamefont
  {K.}~\bibnamefont {Kawata}}, \bibinfo {author} {\bibfnamefont
  {T.}~\bibnamefont {Komatsubara}}, \bibinfo {author} {\bibfnamefont
  {T.}~\bibnamefont {Kubo}}, \bibinfo {author} {\bibfnamefont {K.}~\bibnamefont
  {Kusaka}}, \bibinfo {author} {\bibfnamefont {D.~J.}\ \bibnamefont
  {Morrissey}}, \bibinfo {author} {\bibfnamefont {M.}~\bibnamefont {Ohtake}},
  \emph {et~al.},\ }\href {\doibase 10.1103/PhysRevLett.121.022501} {\bibfield
  {journal} {\bibinfo  {journal} {Physical Review Letters}\ }\textbf {\bibinfo
  {volume} {121}},\ \bibinfo {pages} {022501} (\bibinfo {year}
  {2018})}\BibitemShut {NoStop}%
\bibitem [{\citenamefont {Michimasa}\ \emph {et~al.}(2018)\citenamefont
  {Michimasa}, \citenamefont {Kobayashi}, \citenamefont {Kiyokawa},
  \citenamefont {Ota}, \citenamefont {Ahn}, \citenamefont {Baba}, \citenamefont
  {Berg}, \citenamefont {Dozono}, \citenamefont {Fukuda}, \citenamefont
  {Furuno}, \citenamefont {Ideguchi}, \citenamefont {Inabe}, \citenamefont
  {Kawabata}, \citenamefont {Kawase}, \citenamefont {Kisamori} \emph
  {et~al.}}]{2018Michimasa}%
  \BibitemOpen
  \bibfield  {author} {\bibinfo {author} {\bibfnamefont {S.}~\bibnamefont
  {Michimasa}}, \bibinfo {author} {\bibfnamefont {M.}~\bibnamefont
  {Kobayashi}}, \bibinfo {author} {\bibfnamefont {Y.}~\bibnamefont {Kiyokawa}},
  \bibinfo {author} {\bibfnamefont {S.}~\bibnamefont {Ota}}, \bibinfo {author}
  {\bibfnamefont {D.~S.}\ \bibnamefont {Ahn}}, \bibinfo {author} {\bibfnamefont
  {H.}~\bibnamefont {Baba}}, \bibinfo {author} {\bibfnamefont {G.~P.~A.}\
  \bibnamefont {Berg}}, \bibinfo {author} {\bibfnamefont {M.}~\bibnamefont
  {Dozono}}, \bibinfo {author} {\bibfnamefont {N.}~\bibnamefont {Fukuda}},
  \bibinfo {author} {\bibfnamefont {T.}~\bibnamefont {Furuno}}, \bibinfo
  {author} {\bibfnamefont {E.}~\bibnamefont {Ideguchi}}, \bibinfo {author}
  {\bibfnamefont {N.}~\bibnamefont {Inabe}}, \bibinfo {author} {\bibfnamefont
  {T.}~\bibnamefont {Kawabata}}, \bibinfo {author} {\bibfnamefont
  {S.}~\bibnamefont {Kawase}}, \bibinfo {author} {\bibfnamefont
  {K.}~\bibnamefont {Kisamori}},  \emph {et~al.},\ }\href {\doibase
  10.1103/PhysRevLett.121.022506} {\bibfield  {journal} {\bibinfo  {journal}
  {Physical Review Letters}\ }\textbf {\bibinfo {volume} {121}},\ \bibinfo
  {pages} {022506} (\bibinfo {year} {2018})}\BibitemShut {NoStop}%
\bibitem [{\citenamefont {Wienholtz}\ \emph {et~al.}(2013)\citenamefont
  {Wienholtz}, \citenamefont {Beck}, \citenamefont {Blaum}, \citenamefont
  {Borgmann}, \citenamefont {Breitenfeldt}, \citenamefont {Cakirli},
  \citenamefont {George}, \citenamefont {Herfurth}, \citenamefont {Holt},
  \citenamefont {Kowalska}, \citenamefont {Kreim}, \citenamefont {Lunney},
  \citenamefont {Manea}, \citenamefont {Menéndez}, \citenamefont {Neidherr}
  \emph {et~al.}}]{2013Wienholtz}%
  \BibitemOpen
  \bibfield  {author} {\bibinfo {author} {\bibfnamefont {F.}~\bibnamefont
  {Wienholtz}}, \bibinfo {author} {\bibfnamefont {D.}~\bibnamefont {Beck}},
  \bibinfo {author} {\bibfnamefont {K.}~\bibnamefont {Blaum}}, \bibinfo
  {author} {\bibfnamefont {C.}~\bibnamefont {Borgmann}}, \bibinfo {author}
  {\bibfnamefont {M.}~\bibnamefont {Breitenfeldt}}, \bibinfo {author}
  {\bibfnamefont {R.~B.}\ \bibnamefont {Cakirli}}, \bibinfo {author}
  {\bibfnamefont {S.}~\bibnamefont {George}}, \bibinfo {author} {\bibfnamefont
  {F.}~\bibnamefont {Herfurth}}, \bibinfo {author} {\bibfnamefont {J.~D.}\
  \bibnamefont {Holt}}, \bibinfo {author} {\bibfnamefont {M.}~\bibnamefont
  {Kowalska}}, \bibinfo {author} {\bibfnamefont {S.}~\bibnamefont {Kreim}},
  \bibinfo {author} {\bibfnamefont {D.}~\bibnamefont {Lunney}}, \bibinfo
  {author} {\bibfnamefont {V.}~\bibnamefont {Manea}}, \bibinfo {author}
  {\bibfnamefont {J.}~\bibnamefont {Menéndez}}, \bibinfo {author}
  {\bibfnamefont {D.}~\bibnamefont {Neidherr}},  \emph {et~al.},\ }\href
  {\doibase 10.1038/nature12226} {\bibfield  {journal} {\bibinfo  {journal}
  {Nature}\ }\textbf {\bibinfo {volume} {498}},\ \bibinfo {pages} {346}
  (\bibinfo {year} {2013})}\BibitemShut {NoStop}%
\bibitem [{\citenamefont {Steppenbeck}\ \emph {et~al.}(2013)\citenamefont
  {Steppenbeck}, \citenamefont {Takeuchi}, \citenamefont {Aoi}, \citenamefont
  {Doornenbal}, \citenamefont {Matsushita}, \citenamefont {Wang}, \citenamefont
  {Baba}, \citenamefont {Fukuda}, \citenamefont {Go}, \citenamefont {Honma},
  \citenamefont {Lee}, \citenamefont {Matsui}, \citenamefont {Michimasa},
  \citenamefont {Motobayashi}, \citenamefont {Nishimura} \emph
  {et~al.}}]{2013Steppenbeck}%
  \BibitemOpen
  \bibfield  {author} {\bibinfo {author} {\bibfnamefont {D.}~\bibnamefont
  {Steppenbeck}}, \bibinfo {author} {\bibfnamefont {S.}~\bibnamefont
  {Takeuchi}}, \bibinfo {author} {\bibfnamefont {N.}~\bibnamefont {Aoi}},
  \bibinfo {author} {\bibfnamefont {P.}~\bibnamefont {Doornenbal}}, \bibinfo
  {author} {\bibfnamefont {M.}~\bibnamefont {Matsushita}}, \bibinfo {author}
  {\bibfnamefont {H.}~\bibnamefont {Wang}}, \bibinfo {author} {\bibfnamefont
  {H.}~\bibnamefont {Baba}}, \bibinfo {author} {\bibfnamefont {N.}~\bibnamefont
  {Fukuda}}, \bibinfo {author} {\bibfnamefont {S.}~\bibnamefont {Go}}, \bibinfo
  {author} {\bibfnamefont {M.}~\bibnamefont {Honma}}, \bibinfo {author}
  {\bibfnamefont {J.}~\bibnamefont {Lee}}, \bibinfo {author} {\bibfnamefont
  {K.}~\bibnamefont {Matsui}}, \bibinfo {author} {\bibfnamefont
  {S.}~\bibnamefont {Michimasa}}, \bibinfo {author} {\bibfnamefont
  {T.}~\bibnamefont {Motobayashi}}, \bibinfo {author} {\bibfnamefont
  {D.}~\bibnamefont {Nishimura}},  \emph {et~al.},\ }\href {\doibase
  10.1038/nature12522} {\bibfield  {journal} {\bibinfo  {journal} {Nature}\
  }\textbf {\bibinfo {volume} {502}},\ \bibinfo {pages} {207} (\bibinfo {year}
  {2013})}\BibitemShut {NoStop}%
\bibitem [{\citenamefont {Gade}\ \emph {et~al.}(2006)\citenamefont {Gade},
  \citenamefont {Janssens}, \citenamefont {Bazin}, \citenamefont {Broda},
  \citenamefont {Brown}, \citenamefont {Campbell}, \citenamefont {Carpenter},
  \citenamefont {Cook}, \citenamefont {Deacon}, \citenamefont {Dinca},
  \citenamefont {Fornal}, \citenamefont {Freeman}, \citenamefont {Glasmacher},
  \citenamefont {Hansen}, \citenamefont {Kay} \emph {et~al.}}]{2016Gade}%
  \BibitemOpen
  \bibfield  {author} {\bibinfo {author} {\bibfnamefont {A.}~\bibnamefont
  {Gade}}, \bibinfo {author} {\bibfnamefont {R.~V}~\bibnamefont {Janssens}},
  \bibinfo {author} {\bibfnamefont {D.}~\bibnamefont {Bazin}}, \bibinfo
  {author} {\bibfnamefont {R.}~\bibnamefont {Broda}}, \bibinfo {author}
  {\bibfnamefont {B.~A.}\ \bibnamefont {Brown}}, \bibinfo {author}
  {\bibfnamefont {C.~M.}\ \bibnamefont {Campbell}}, \bibinfo {author}
  {\bibfnamefont {M.~P.}\ \bibnamefont {Carpenter}}, \bibinfo {author}
  {\bibfnamefont {J.~M.}~\bibnamefont {Cook}}, \bibinfo {author} {\bibfnamefont
  {A.~N.}~\bibnamefont {Deacon}}, \bibinfo {author} {\bibfnamefont {D.-C.}\
  \bibnamefont {Dinca}}, \bibinfo {author} {\bibfnamefont {B.}~\bibnamefont
  {Fornal}}, \bibinfo {author} {\bibfnamefont {S.~J.}~\bibnamefont {Freeman}},
  \bibinfo {author} {\bibfnamefont {T.}~\bibnamefont {Glasmacher}}, \bibinfo
  {author} {\bibfnamefont {P.~G.}~\bibnamefont {Hansen}}, \bibinfo {author}
  {\bibfnamefont {B.~P.}\ \bibnamefont {Kay}},  \emph {et~al.},\ }\href
  {\doibase 10.1103/PhysRevC.74.021302} {\bibfield  {journal} {\bibinfo
  {journal} {Physical Review C}\ }\textbf {\bibinfo {volume} {74}},\ \bibinfo
  {pages} {021302(R)} (\bibinfo {year} {2006})}\BibitemShut {NoStop}%
\bibitem [{\citenamefont {Garcia~Ruiz}\ \emph {et~al.}(2016)\citenamefont
  {Garcia~Ruiz}, \citenamefont {Bissell}, \citenamefont {Blaum}, \citenamefont
  {Ekstr{\"o}m}, \citenamefont {Fr{\"o}mmgen}, \citenamefont {Hagen},
  \citenamefont {Hammen}, \citenamefont {Hebeler}, \citenamefont {Holt},
  \citenamefont {Jansen}, \citenamefont {Kowalska}, \citenamefont {Kreim},
  \citenamefont {Nazarewicz}, \citenamefont {Neugart}, \citenamefont {Neyens}
  \emph {et~al.}}]{2016Ruiz}%
  \BibitemOpen
  \bibfield  {author} {\bibinfo {author} {\bibfnamefont {R.~F.}\ \bibnamefont
  {Garcia~Ruiz}}, \bibinfo {author} {\bibfnamefont {M.~L.}\ \bibnamefont
  {Bissell}}, \bibinfo {author} {\bibfnamefont {K.}~\bibnamefont {Blaum}},
  \bibinfo {author} {\bibfnamefont {A.}~\bibnamefont {Ekstr{\"o}m}}, \bibinfo
  {author} {\bibfnamefont {N.}~\bibnamefont {Fr{\"o}mmgen}}, \bibinfo {author}
  {\bibfnamefont {G.}~\bibnamefont {Hagen}}, \bibinfo {author} {\bibfnamefont
  {M.}~\bibnamefont {Hammen}}, \bibinfo {author} {\bibfnamefont
  {K.}~\bibnamefont {Hebeler}}, \bibinfo {author} {\bibfnamefont {J.~D.}\
  \bibnamefont {Holt}}, \bibinfo {author} {\bibfnamefont {G.~R.}\ \bibnamefont
  {Jansen}}, \bibinfo {author} {\bibfnamefont {M.}~\bibnamefont {Kowalska}},
  \bibinfo {author} {\bibfnamefont {K.}~\bibnamefont {Kreim}}, \bibinfo
  {author} {\bibfnamefont {W.}~\bibnamefont {Nazarewicz}}, \bibinfo {author}
  {\bibfnamefont {R.}~\bibnamefont {Neugart}}, \bibinfo {author} {\bibfnamefont
  {G.}~\bibnamefont {Neyens}},  \emph {et~al.},\ }\href {\doibase
  10.1038/nphys3645} {\bibfield  {journal} {\bibinfo  {journal} {Nature}\
  }\textbf {\bibinfo {volume} {12}},\ \bibinfo {pages} {594–598} (\bibinfo
  {year} {2016})}\BibitemShut {NoStop}%
\bibitem [{\citenamefont {Tanaka}\ \emph {et~al.}(2020)\citenamefont {Tanaka},
  \citenamefont {Takechi}, \citenamefont {Homma}, \citenamefont {Fukuda},
  \citenamefont {Nishimura}, \citenamefont {Suzuki}, \citenamefont {Tanaka},
  \citenamefont {Moriguchi}, \citenamefont {Ahn}, \citenamefont {Aimaganbetov},
  \citenamefont {Amano}, \citenamefont {Arakawa}, \citenamefont {Bagchi},
  \citenamefont {Behr}, \citenamefont {Burtebayev} \emph
  {et~al.}}]{2019Tanaka}%
  \BibitemOpen
  \bibfield  {author} {\bibinfo {author} {\bibfnamefont {M.}~\bibnamefont
  {Tanaka}}, \bibinfo {author} {\bibfnamefont {M.}~\bibnamefont {Takechi}},
  \bibinfo {author} {\bibfnamefont {A.}~\bibnamefont {Homma}}, \bibinfo
  {author} {\bibfnamefont {M.}~\bibnamefont {Fukuda}}, \bibinfo {author}
  {\bibfnamefont {D.}~\bibnamefont {Nishimura}}, \bibinfo {author}
  {\bibfnamefont {T.}~\bibnamefont {Suzuki}}, \bibinfo {author} {\bibfnamefont
  {Y.}~\bibnamefont {Tanaka}}, \bibinfo {author} {\bibfnamefont
  {T.}~\bibnamefont {Moriguchi}}, \bibinfo {author} {\bibfnamefont
  {D.}~\bibnamefont {Ahn}}, \bibinfo {author} {\bibfnamefont {A.}~\bibnamefont
  {Aimaganbetov}}, \bibinfo {author} {\bibfnamefont {M.}~\bibnamefont {Amano}},
  \bibinfo {author} {\bibfnamefont {H.}~\bibnamefont {Arakawa}}, \bibinfo
  {author} {\bibfnamefont {S.}~\bibnamefont {Bagchi}}, \bibinfo {author}
  {\bibfnamefont {K.-H.}\ \bibnamefont {Behr}}, \bibinfo {author}
  {\bibfnamefont {N.}~\bibnamefont {Burtebayev}},  \emph {et~al.},\ }\href
  {\doibase 10.1103/PhysRevLett.124.102501} {\bibfield  {journal} {\bibinfo
  {journal} {Physical Review Letters}\ }\textbf {\bibinfo {volume} {124}},\
  \bibinfo {pages} {102501} (\bibinfo {year} {2020})}\BibitemShut {NoStop}%
\bibitem [{\citenamefont {Liu}\ \emph {et~al.}(2019)\citenamefont {Liu},
  \citenamefont {Obertelli}, \citenamefont {Doornenbal}, \citenamefont
  {Bertulani}, \citenamefont {Hagen}, \citenamefont {Holt}, \citenamefont
  {Jansen}, \citenamefont {Morris}, \citenamefont {Schwenk}, \citenamefont
  {Stroberg}, \citenamefont {Achouri}, \citenamefont {Baba}, \citenamefont
  {Browne}, \citenamefont {Calvet}, \citenamefont {Ch\^ateau} \emph
  {et~al.}}]{2019Liu}%
  \BibitemOpen
  \bibfield  {author} {\bibinfo {author} {\bibfnamefont {H.~N.}\ \bibnamefont
  {Liu}}, \bibinfo {author} {\bibfnamefont {A.}~\bibnamefont {Obertelli}},
  \bibinfo {author} {\bibfnamefont {P.}~\bibnamefont {Doornenbal}}, \bibinfo
  {author} {\bibfnamefont {C.~A.}\ \bibnamefont {Bertulani}}, \bibinfo {author}
  {\bibfnamefont {G.}~\bibnamefont {Hagen}}, \bibinfo {author} {\bibfnamefont
  {J.~D.}\ \bibnamefont {Holt}}, \bibinfo {author} {\bibfnamefont {G.~R.}\
  \bibnamefont {Jansen}}, \bibinfo {author} {\bibfnamefont {T.~D.}\
  \bibnamefont {Morris}}, \bibinfo {author} {\bibfnamefont {A.}~\bibnamefont
  {Schwenk}}, \bibinfo {author} {\bibfnamefont {R.}~\bibnamefont {Stroberg}},
  \bibinfo {author} {\bibfnamefont {N.}~\bibnamefont {Achouri}}, \bibinfo
  {author} {\bibfnamefont {H.}~\bibnamefont {Baba}}, \bibinfo {author}
  {\bibfnamefont {F.}~\bibnamefont {Browne}}, \bibinfo {author} {\bibfnamefont
  {D.}~\bibnamefont {Calvet}}, \bibinfo {author} {\bibfnamefont
  {F.}~\bibnamefont {Ch\^ateau}},  \emph {et~al.},\ }\href {\doibase
  10.1103/PhysRevLett.122.072502} {\bibfield  {journal} {\bibinfo  {journal}
  {Physical Review Letters}\ }\textbf {\bibinfo {volume} {122}},\ \bibinfo
  {pages} {072502} (\bibinfo {year} {2019})}\BibitemShut {NoStop}%
\bibitem [{\citenamefont {Leistenschneider}\ \emph {et~al.}(2020)\citenamefont
  {Leistenschneider}, \citenamefont {Dunling}, \citenamefont {Bollen},
  \citenamefont {Brown}, \citenamefont {Dilling}, \citenamefont {Hamaker},
  \citenamefont {Holt}, \citenamefont {Kwiatkowski}, \citenamefont {Miyagi},
  \citenamefont {Porter}, \citenamefont {Puentes}, \citenamefont {Redshaw},
  \citenamefont {Reiter}, \citenamefont {Ringle}, \citenamefont {Sandler} \emph
  {et~al.}}]{2020Leistenschneider}%
  \BibitemOpen
  \bibfield  {author} {\bibinfo {author} {\bibfnamefont {E.}~\bibnamefont
  {Leistenschneider}}, \bibinfo {author} {\bibfnamefont {E.}~\bibnamefont
  {Dunling}}, \bibinfo {author} {\bibfnamefont {G.}~\bibnamefont {Bollen}},
  \bibinfo {author} {\bibfnamefont {B.}~\bibnamefont {Brown}}, \bibinfo
  {author} {\bibfnamefont {J.}~\bibnamefont {Dilling}}, \bibinfo {author}
  {\bibfnamefont {A.}~\bibnamefont {Hamaker}}, \bibinfo {author} {\bibfnamefont
  {J.}~\bibnamefont {Holt}}, \bibinfo {author} {\bibfnamefont {A.~A.}\
  \bibnamefont {Kwiatkowski}}, \bibinfo {author} {\bibfnamefont
  {T.}~\bibnamefont {Miyagi}}, \bibinfo {author} {\bibfnamefont {W.~S.}\
  \bibnamefont {Porter}}, \bibinfo {author} {\bibfnamefont {D.}~\bibnamefont
  {Puentes}}, \bibinfo {author} {\bibfnamefont {M.}~\bibnamefont {Redshaw}},
  \bibinfo {author} {\bibfnamefont {M.~P.}\ \bibnamefont {Reiter}}, \bibinfo
  {author} {\bibfnamefont {R.}~\bibnamefont {Ringle}}, \bibinfo {author}
  {\bibfnamefont {R.}~\bibnamefont {Sandler}},  \emph {et~al.},\ }\href@noop {}
  {\  (\bibinfo {year} {2020})},\ \Eprint {http://arxiv.org/abs/2006.01302}
  {arXiv:2006.01302 [nucl-ex]} \BibitemShut {NoStop}%
\bibitem [{\citenamefont {Cortés}\ \emph {et~al.}(2020)\citenamefont
  {Cortés}, \citenamefont {Rodriguez}, \citenamefont {Doornenbal},
  \citenamefont {Obertelli}, \citenamefont {Holt}, \citenamefont {Lenzi},
  \citenamefont {Menéndez}, \citenamefont {Nowacki}, \citenamefont {Ogata},
  \citenamefont {Poves}, \citenamefont {Rodríguez}, \citenamefont {Schwenk},
  \citenamefont {Simonis}, \citenamefont {Stroberg}, \citenamefont {Yoshida}
  \emph {et~al.}}]{2020Cortes}%
  \BibitemOpen
  \bibfield  {author} {\bibinfo {author} {\bibfnamefont {M.}~\bibnamefont
  {Cortés}}, \bibinfo {author} {\bibfnamefont {W.}~\bibnamefont {Rodriguez}},
  \bibinfo {author} {\bibfnamefont {P.}~\bibnamefont {Doornenbal}}, \bibinfo
  {author} {\bibfnamefont {A.}~\bibnamefont {Obertelli}}, \bibinfo {author}
  {\bibfnamefont {J.}~\bibnamefont {Holt}}, \bibinfo {author} {\bibfnamefont
  {S.}~\bibnamefont {Lenzi}}, \bibinfo {author} {\bibfnamefont
  {J.}~\bibnamefont {Menéndez}}, \bibinfo {author} {\bibfnamefont
  {F.}~\bibnamefont {Nowacki}}, \bibinfo {author} {\bibfnamefont
  {K.}~\bibnamefont {Ogata}}, \bibinfo {author} {\bibfnamefont
  {A.}~\bibnamefont {Poves}}, \bibinfo {author} {\bibfnamefont
  {T.}~\bibnamefont {Rodríguez}}, \bibinfo {author} {\bibfnamefont
  {A.}~\bibnamefont {Schwenk}}, \bibinfo {author} {\bibfnamefont
  {J.}~\bibnamefont {Simonis}}, \bibinfo {author} {\bibfnamefont
  {S.}~\bibnamefont {Stroberg}}, \bibinfo {author} {\bibfnamefont
  {K.}~\bibnamefont {Yoshida}},  \emph {et~al.},\ }\href {\doibase
  10.1016/j.physletb.2019.135071} {\bibfield  {journal} {\bibinfo  {journal}
  {Physics Letters B}\ }\textbf {\bibinfo {volume} {800}},\ \bibinfo {pages}
  {135071} (\bibinfo {year} {2020})}\BibitemShut {NoStop}%
\bibitem [{\citenamefont {NuPECC}(2017)}]{2017nupecc}%
  \BibitemOpen
  \bibfield  {author} {\bibinfo {author} {\bibnamefont {NuPECC}},\ }\href
  {\doibase 10.1051/epn/2017403} {\bibfield  {journal} {\bibinfo  {journal}
  {Long Range Plan 2017. Perspectives in Nuclear Physics}\ } (\bibinfo {year}
  {2017}),\ 10.1051/epn/2017403}\BibitemShut {NoStop}%
\bibitem [{201(2018)}]{2018frib}%
  \BibitemOpen
  \href {\doibase 10.1088/1361-6471/ab26cc} {\bibfield  {journal} {\bibinfo
  {journal} {Edited by G. W. Severin. Isotope harvesting at FRIB: Additional
  opportunities for scientific discovery}\ } (\bibinfo {year} {2018}),\
  10.1088/1361-6471/ab26cc}\BibitemShut {NoStop}%
\bibitem [{\citenamefont {Hagen}\ \emph {et~al.}(2012)\citenamefont {Hagen},
  \citenamefont {Hjorth-Jensen}, \citenamefont {Jansen}, \citenamefont
  {Machleidt},\ and\ \citenamefont {Papenbrock}}]{2012Hagen}%
  \BibitemOpen
  \bibfield  {author} {\bibinfo {author} {\bibfnamefont {G.}~\bibnamefont
  {Hagen}}, \bibinfo {author} {\bibfnamefont {M.}~\bibnamefont
  {Hjorth-Jensen}}, \bibinfo {author} {\bibfnamefont {G.~R.}\ \bibnamefont
  {Jansen}}, \bibinfo {author} {\bibfnamefont {R.}~\bibnamefont {Machleidt}}, \
  and\ \bibinfo {author} {\bibfnamefont {T.}~\bibnamefont {Papenbrock}},\
  }\href {\doibase 10.1103/PhysRevLett.109.032502} {\bibfield  {journal}
  {\bibinfo  {journal} {Physical Review Letters}\ }\textbf {\bibinfo {volume}
  {109}},\ \bibinfo {pages} {032502} (\bibinfo {year} {2012})}\BibitemShut
  {NoStop}%
\bibitem [{\citenamefont {Tichai}\ \emph {et~al.}(2018)\citenamefont {Tichai},
  \citenamefont {Arthuis}, \citenamefont {Duguet}, \citenamefont {Hergert},
  \citenamefont {Somà},\ and\ \citenamefont {Roth}}]{2018Tichai}%
  \BibitemOpen
  \bibfield  {author} {\bibinfo {author} {\bibfnamefont {A.}~\bibnamefont
  {Tichai}}, \bibinfo {author} {\bibfnamefont {P.}~\bibnamefont {Arthuis}},
  \bibinfo {author} {\bibfnamefont {T.}~\bibnamefont {Duguet}}, \bibinfo
  {author} {\bibfnamefont {H.}~\bibnamefont {Hergert}}, \bibinfo {author}
  {\bibfnamefont {V.}~\bibnamefont {Somà}}, \ and\ \bibinfo {author}
  {\bibfnamefont {R.}~\bibnamefont {Roth}},\ }\href {\doibase
  10.1016/j.physletb.2018.09.044} {\bibfield  {journal} {\bibinfo  {journal}
  {Physics Letters B}\ }\textbf {\bibinfo {volume} {786}},\ \bibinfo {pages}
  {195} (\bibinfo {year} {2018})}\BibitemShut {NoStop}%
\bibitem [{\citenamefont {Soma}(2020)}]{2020Soma}%
  \BibitemOpen
  \bibfield  {author} {\bibinfo {author} {\bibfnamefont {V.}~\bibnamefont
  {Soma}},\ }\href@noop {} {\  (\bibinfo {year} {2020})},\ \Eprint
  {http://arxiv.org/abs/2003.11321} {arXiv:2003.11321 [nucl-th]} \BibitemShut
  {NoStop}%
\bibitem [{\citenamefont {Agbemava}\ \emph {et~al.}(2014)\citenamefont
  {Agbemava}, \citenamefont {Afanasjev}, \citenamefont {Ray},\ and\
  \citenamefont {Ring}}]{2014Agbemava}%
  \BibitemOpen
  \bibfield  {author} {\bibinfo {author} {\bibfnamefont {S.~E.}\ \bibnamefont
  {Agbemava}}, \bibinfo {author} {\bibfnamefont {A.~V.}\ \bibnamefont
  {Afanasjev}}, \bibinfo {author} {\bibfnamefont {D.}~\bibnamefont {Ray}}, \
  and\ \bibinfo {author} {\bibfnamefont {P.}~\bibnamefont {Ring}},\ }\href
  {\doibase 10.1103/PhysRevC.89.054320} {\bibfield  {journal} {\bibinfo
  {journal} {Physical Review C}\ }\textbf {\bibinfo {volume} {89}},\ \bibinfo
  {pages} {054320} (\bibinfo {year} {2014})}\BibitemShut {NoStop}%
\bibitem [{\citenamefont {Neufcourt}\ \emph {et~al.}(2019)\citenamefont
  {Neufcourt}, \citenamefont {Cao}, \citenamefont {Nazarewicz}, \citenamefont
  {Olsen},\ and\ \citenamefont {Viens}}]{2019Neufcourt}%
  \BibitemOpen
  \bibfield  {author} {\bibinfo {author} {\bibfnamefont {L.}~\bibnamefont
  {Neufcourt}}, \bibinfo {author} {\bibfnamefont {Y.}~\bibnamefont {Cao}},
  \bibinfo {author} {\bibfnamefont {W.}~\bibnamefont {Nazarewicz}}, \bibinfo
  {author} {\bibfnamefont {E.}~\bibnamefont {Olsen}}, \ and\ \bibinfo {author}
  {\bibfnamefont {F.}~\bibnamefont {Viens}},\ }\href {\doibase
  10.1103/PhysRevLett.122.062502} {\bibfield  {journal} {\bibinfo  {journal}
  {Physical Review Letters}\ }\textbf {\bibinfo {volume} {122}},\ \bibinfo
  {pages} {062502} (\bibinfo {year} {2019})}\BibitemShut {NoStop}%
\bibitem [{\citenamefont {Li}\ \emph {et~al.}(2020)\citenamefont {Li},
  \citenamefont {Hu}, \citenamefont {Wu}, \citenamefont {Gao}, \citenamefont
  {Dai},\ and\ \citenamefont {Xu}}]{2020Li}%
  \BibitemOpen
  \bibfield  {author} {\bibinfo {author} {\bibfnamefont {J.~G.}\ \bibnamefont
  {Li}}, \bibinfo {author} {\bibfnamefont {B.~S.}\ \bibnamefont {Hu}}, \bibinfo
  {author} {\bibfnamefont {Q.}~\bibnamefont {Wu}}, \bibinfo {author}
  {\bibfnamefont {Y.}~\bibnamefont {Gao}}, \bibinfo {author} {\bibfnamefont
  {S.~J.}\ \bibnamefont {Dai}}, \ and\ \bibinfo {author} {\bibfnamefont
  {F.~R.}\ \bibnamefont {Xu}},\ }\href@noop {} {\bibfield  {journal} {\bibinfo
  {journal} {Physical Review C}\ }\textbf {\bibinfo {volume} {102}},\ \bibinfo
  {pages} {034302} (\bibinfo {year} {2020})}\BibitemShut {NoStop}%
\bibitem [{\citenamefont {Lane}(1964)}]{1964Lane}%
  \BibitemOpen
  \bibfield  {author} {\bibinfo {author} {\bibfnamefont {A.~M.}\ \bibnamefont
  {Lane}},\ }\href@noop {} {\emph {\bibinfo {title} {Nuclear Theory, Pairing
  Force Correlations and Collective Motion}}}\ (\bibinfo  {publisher}
  {Benjamin, New York},\ \bibinfo {year} {1964})\BibitemShut {NoStop}%
\bibitem [{\citenamefont {Broglia}\ and\ \citenamefont
  {Zelevinsky}(2013)}]{2013Zelevinsky}%
  \BibitemOpen
  \bibfield  {author} {\bibinfo {author} {\bibfnamefont {R.}~\bibnamefont
  {Broglia}}\ and\ \bibinfo {author} {\bibfnamefont {V.}~\bibnamefont
  {Zelevinsky}},\ }\href@noop {} {\emph {\bibinfo {title} {Fifty Years of
  Nuclear BCS. Pairing in Finite Systems}}}\ (\bibinfo  {publisher} {World
  Scientific. Singapore},\ \bibinfo {year} {2013})\BibitemShut {NoStop}%
\bibitem [{\citenamefont {Gorkov}(1958)}]{1958Gorkov}%
  \BibitemOpen
  \bibfield  {author} {\bibinfo {author} {\bibfnamefont {L.~P.}\ \bibnamefont
  {Gorkov}},\ }\href@noop {} {\bibfield  {journal} {\bibinfo  {journal} {Soviet
  Phys. JETP}\ }\textbf {\bibinfo {volume} {34(7)}},\ \bibinfo {pages} {505}
  (\bibinfo {year} {1958})}\BibitemShut {NoStop}%
\bibitem [{\citenamefont {Gorkov}(2010)}]{2011Gorkov}%
  \BibitemOpen
  \bibfield  {author} {\bibinfo {author} {\bibfnamefont {L.~P.}\ \bibnamefont
  {Gorkov}},\ }\href {\doibase 10.1142/S0217979210056372} {\bibfield  {journal}
  {\bibinfo  {journal} {International Journal of Modern Physics B}\ }\textbf
  {\bibinfo {volume} {24}},\ \bibinfo {pages} {3835} (\bibinfo {year}
  {2010})}\BibitemShut {NoStop}%
\bibitem [{\citenamefont {Som\`a}\ \emph {et~al.}(2011)\citenamefont {Som\`a},
  \citenamefont {Duguet},\ and\ \citenamefont {Barbieri}}]{2011Soma}%
  \BibitemOpen
  \bibfield  {author} {\bibinfo {author} {\bibfnamefont {V.}~\bibnamefont
  {Som\`a}}, \bibinfo {author} {\bibfnamefont {T.}~\bibnamefont {Duguet}}, \
  and\ \bibinfo {author} {\bibfnamefont {C.}~\bibnamefont {Barbieri}},\ }\href
  {\doibase 10.1103/PhysRevC.84.064317} {\bibfield  {journal} {\bibinfo
  {journal} {Phys. Rev. C}\ }\textbf {\bibinfo {volume} {84}},\ \bibinfo
  {pages} {064317} (\bibinfo {year} {2011})}\BibitemShut {NoStop}%
\bibitem [{\citenamefont {Som\`a}\ \emph {et~al.}(2014)\citenamefont {Som\`a},
  \citenamefont {Cipollone}, \citenamefont {Barbieri}, \citenamefont
  {Navr\'atil},\ and\ \citenamefont {Duguet}}]{2014Soma}%
  \BibitemOpen
  \bibfield  {author} {\bibinfo {author} {\bibfnamefont {V.}~\bibnamefont
  {Som\`a}}, \bibinfo {author} {\bibfnamefont {A.}~\bibnamefont {Cipollone}},
  \bibinfo {author} {\bibfnamefont {C.}~\bibnamefont {Barbieri}}, \bibinfo
  {author} {\bibfnamefont {P.}~\bibnamefont {Navr\'atil}}, \ and\ \bibinfo
  {author} {\bibfnamefont {T.}~\bibnamefont {Duguet}},\ }\href {\doibase
  10.1103/PhysRevC.89.061301} {\bibfield  {journal} {\bibinfo  {journal} {Phys.
  Rev. C}\ }\textbf {\bibinfo {volume} {89}},\ \bibinfo {pages} {061301(R)}
  (\bibinfo {year} {2014})}\BibitemShut {NoStop}%
\bibitem [{\citenamefont {Som\`a}\ \emph {et~al.}(2013)\citenamefont {Som\`a},
  \citenamefont {Barbieri},\ and\ \citenamefont {Duguet}}]{2013Soma}%
  \BibitemOpen
  \bibfield  {author} {\bibinfo {author} {\bibfnamefont {V.}~\bibnamefont
  {Som\`a}}, \bibinfo {author} {\bibfnamefont {C.}~\bibnamefont {Barbieri}}, \
  and\ \bibinfo {author} {\bibfnamefont {T.}~\bibnamefont {Duguet}},\ }\href
  {\doibase 10.1103/PhysRevC.87.011303} {\bibfield  {journal} {\bibinfo
  {journal} {Phys. Rev. C}\ }\textbf {\bibinfo {volume} {87}},\ \bibinfo
  {pages} {011303(R)} (\bibinfo {year} {2013})}\BibitemShut {NoStop}%
\bibitem [{\citenamefont {Id~Betan}\ and\ \citenamefont
  {Repetto}(2020)}]{2020IdBetan}%
  \BibitemOpen
  \bibfield  {author} {\bibinfo {author} {\bibfnamefont {R.~M.}\ \bibnamefont
  {Id~Betan}}\ and\ \bibinfo {author} {\bibfnamefont {C.~E.}\ \bibnamefont
  {Repetto}},\ }\href {\doibase 10.1016/j.nuclphysa.2019.121676} {\bibfield
  {journal} {\bibinfo  {journal} {Nuclear Physics A}\ }\textbf {\bibinfo
  {volume} {994}},\ \bibinfo {pages} {121676} (\bibinfo {year}
  {2020})}\BibitemShut {NoStop}%
\bibitem [{\citenamefont {Richardson}(1963)}]{Richardson1963}%
  \BibitemOpen
  \bibfield  {author} {\bibinfo {author} {\bibfnamefont {R.~W.}\ \bibnamefont
  {Richardson}},\ }\href {\doibase 10.1016/0031-9163(63)90259-2} {\bibfield
  {journal} {\bibinfo  {journal} {Physical Review Letters}\ }\textbf {\bibinfo
  {volume} {3}},\ \bibinfo {pages} {277 } (\bibinfo {year} {1963})}\BibitemShut
  {NoStop}%
\bibitem [{\citenamefont {Richardson}\ and\ \citenamefont
  {Sherman}(1964{\natexlab{a}})}]{1964Richardson}%
  \BibitemOpen
  \bibfield  {author} {\bibinfo {author} {\bibfnamefont {R.~W.}\ \bibnamefont
  {Richardson}}\ and\ \bibinfo {author} {\bibfnamefont {N.}~\bibnamefont
  {Sherman}},\ }\href {\doibase 10.1016/0029-5582(64)90687-X} {\bibfield
  {journal} {\bibinfo  {journal} {Nuclear Physics}\ }\textbf {\bibinfo
  {volume} {52}},\ \bibinfo {pages} {221} (\bibinfo {year}
  {1964}{\natexlab{a}})}\BibitemShut {NoStop}%
\bibitem [{\citenamefont {{Id~Betan}}(2012{\natexlab{a}})}]{IdBetan2012}%
  \BibitemOpen
  \bibfield  {author} {\bibinfo {author} {\bibfnamefont {R.}~\bibnamefont {{Id~Betan}}},\ }\href {\doibase 10.1103/PhysRevC.85.064309} {\bibfield  {journal}
  {\bibinfo  {journal} {Physical Review C - Nuclear Physics}\ }\textbf
  {\bibinfo {volume} {85}},\ \bibinfo {pages} {064309} (\bibinfo {year}
  {2012}{\natexlab{a}})}\BibitemShut {NoStop}%
\bibitem [{\citenamefont {{Id Betan}}(2012{\natexlab{b}})}]{IdBetan2012P}%
  \BibitemOpen
  \bibfield  {author} {\bibinfo {author} {\bibfnamefont {R.}~\bibnamefont {{Id~Betan}}},\ }\href {\doibase 10.1016/j.nuclphysa.2012.01.026} {\bibfield
  {journal} {\bibinfo  {journal} {Nuclear Physics A}\ }\textbf {\bibinfo
  {volume} {879}},\ \bibinfo {pages} {14} (\bibinfo {year}
  {2012}{\natexlab{b}})}\BibitemShut {NoStop}%
\bibitem [{\citenamefont {Id~Betan}(2017)}]{2017IdBetan}%
  \BibitemOpen
  \bibfield  {author} {\bibinfo {author} {\bibfnamefont {R.~M.}\ \bibnamefont
  {Id~Betan}},\ }\href {\doibase 10.1088/1742-6596/839/1/012003} {\bibfield
  {journal} {\bibinfo  {journal} {IOP Conf. Series: Journal of Physics}\
  }\textbf {\bibinfo {volume} {839}},\ \bibinfo {pages} {012003} (\bibinfo
  {year} {2017})}\BibitemShut {NoStop}%
\bibitem [{\citenamefont {Beth}\ and\ \citenamefont
  {Uhlenbeck}(1937)}]{Beth1937II}%
  \BibitemOpen
  \bibfield  {author} {\bibinfo {author} {\bibfnamefont {E.}~\bibnamefont
  {Beth}}\ and\ \bibinfo {author} {\bibfnamefont {G.~E.}\ \bibnamefont
  {Uhlenbeck}},\ }\href {\doibase 10.1016/S0031-8914(37)80189-5} {\bibfield
  {journal} {\bibinfo  {journal} {Physica}\ }\textbf {\bibinfo {volume} {4}},\
  \bibinfo {pages} {915} (\bibinfo {year} {1937})}\BibitemShut {NoStop}%
\bibitem [{\citenamefont {Pittel}(2015)}]{Pittel2015}%
  \BibitemOpen
  \bibfield  {author} {\bibinfo {author} {\bibfnamefont {S.}~\bibnamefont
  {Pittel}},\ }\href {\doibase 10.1088/1742-6596/578/1/012011} {\bibfield
  {journal} {\bibinfo  {journal} {Journal of Physics: Conference Series}\
  }\textbf {\bibinfo {volume} {578}},\ \bibinfo {pages} {8} (\bibinfo {year}
  {2015})}\BibitemShut {NoStop}%
\bibitem [{\citenamefont {von Delft}\ and\ \citenamefont
  {Braun}(1999)}]{VonDelft1999}%
  \BibitemOpen
  \bibfield  {author} {\bibinfo {author} {\bibfnamefont {J.}~\bibnamefont {von
  Delft}}\ and\ \bibinfo {author} {\bibfnamefont {F.}~\bibnamefont {Braun}},\
  }\href@noop {} {\  (\bibinfo {year} {1999})},\ \Eprint
  {http://arxiv.org/abs/cond-mat/9911058} {arXiv:cond-mat/9911058
  [cond-mat.str-el]} \BibitemShut {NoStop}%
\bibitem [{\citenamefont {Nilsson}\ \emph {et~al.}(1969)\citenamefont
  {Nilsson}, \citenamefont {Tsang}, \citenamefont {Sobiczewski}, \citenamefont
  {Szymański}, \citenamefont {Wycech}, \citenamefont {Gustafson},
  \citenamefont {Lamm}, \citenamefont {Möller},\ and\ \citenamefont
  {Nilsson}}]{1969Nilsson}%
  \BibitemOpen
  \bibfield  {author} {\bibinfo {author} {\bibfnamefont {S.~G.}\ \bibnamefont
  {Nilsson}}, \bibinfo {author} {\bibfnamefont {C.~F.}\ \bibnamefont {Tsang}},
  \bibinfo {author} {\bibfnamefont {A.}~\bibnamefont {Sobiczewski}}, \bibinfo
  {author} {\bibfnamefont {Z.}~\bibnamefont {Szymański}}, \bibinfo {author}
  {\bibfnamefont {S.}~\bibnamefont {Wycech}}, \bibinfo {author} {\bibfnamefont
  {C.}~\bibnamefont {Gustafson}}, \bibinfo {author} {\bibfnamefont {I.-L.}\
  \bibnamefont {Lamm}}, \bibinfo {author} {\bibfnamefont {P.}~\bibnamefont
  {Möller}}, \ and\ \bibinfo {author} {\bibfnamefont {B.}~\bibnamefont
  {Nilsson}},\ }\href {\doibase 10.1016/0375-9474(69)90809-4} {\bibfield
  {journal} {\bibinfo  {journal} {Nuclear Physics A}\ }\textbf {\bibinfo
  {volume} {131}},\ \bibinfo {pages} {1 } (\bibinfo {year} {1969})}\BibitemShut
  {NoStop}%
\bibitem [{\citenamefont {{Id Betan}}\ and\ \citenamefont
  {Repetto}(2017)}]{IdBetan2017-R}%
  \BibitemOpen
  \bibfield  {author} {\bibinfo {author} {\bibfnamefont {R.~M.}\ \bibnamefont
  {{Id Betan}}}\ and\ \bibinfo {author} {\bibfnamefont {C.~E.}\ \bibnamefont
  {Repetto}},\ }\href {\doibase 10.1016/j.nuclphysa.2017.02.001} {\bibfield
  {journal} {\bibinfo  {journal} {Nuclear Physics A}\ }\textbf {\bibinfo
  {volume} {960}},\ \bibinfo {pages} {131} (\bibinfo {year}
  {2017})}\BibitemShut {NoStop}%
\bibitem [{\citenamefont {Beiner}\ and\ \citenamefont
  {Lombard}(1975)}]{Beiner1975}%
  \BibitemOpen
  \bibfield  {author} {\bibinfo {author} {\bibfnamefont {M.}~\bibnamefont
  {Beiner}}\ and\ \bibinfo {author} {\bibfnamefont {R.~J.}\ \bibnamefont
  {Lombard}},\ }\href {\doibase https://doi.org/10.1016/0375-9474(75)90089-5}
  {\bibfield  {journal} {\bibinfo  {journal} {Nuclear Physics A}\ }\textbf
  {\bibinfo {volume} {249}},\ \bibinfo {pages} {1} (\bibinfo {year}
  {1975})}\BibitemShut {NoStop}%
\bibitem [{\citenamefont {Dobaczewski}\ \emph {et~al.}(1996)\citenamefont
  {Dobaczewski}, \citenamefont {Nazarewicz}, \citenamefont {Werner},
  \citenamefont {Berger}, \citenamefont {Chinn},\ and\ \citenamefont
  {Decharg\'e}}]{1996Dobaczewski}%
  \BibitemOpen
  \bibfield  {author} {\bibinfo {author} {\bibfnamefont {J.}~\bibnamefont
  {Dobaczewski}}, \bibinfo {author} {\bibfnamefont {W.}~\bibnamefont
  {Nazarewicz}}, \bibinfo {author} {\bibfnamefont {T.~R.}\ \bibnamefont
  {Werner}}, \bibinfo {author} {\bibfnamefont {J.~F.}\ \bibnamefont {Berger}},
  \bibinfo {author} {\bibfnamefont {C.~R.}\ \bibnamefont {Chinn}}, \ and\
  \bibinfo {author} {\bibfnamefont {J.}~\bibnamefont {Decharg\'e}},\ }\href
  {\doibase 10.1103/PhysRevC.53.2809} {\bibfield  {journal} {\bibinfo
  {journal} {Physical Review C}\ }\textbf {\bibinfo {volume} {53}},\ \bibinfo
  {pages} {2809} (\bibinfo {year} {1996})}\BibitemShut {NoStop}%
\bibitem [{\citenamefont {Zenihiro}\ \emph {et~al.}(2018)\citenamefont
  {Zenihiro}, \citenamefont {Sakaguchi}, \citenamefont {Terashima},
  \citenamefont {Uesaka}, \citenamefont {Hagen}, \citenamefont {Itoh},
  \citenamefont {Murakami}, \citenamefont {Nakatsugawa}, \citenamefont
  {Ohnishi}, \citenamefont {Sagawa}, \citenamefont {Takeda}, \citenamefont
  {Uchida}, \citenamefont {Yoshida}, \citenamefont {Yoshida},\ and\
  \citenamefont {Yosoi}}]{2018Zenihiro}%
  \BibitemOpen
  \bibfield  {author} {\bibinfo {author} {\bibfnamefont {J.}~\bibnamefont
  {Zenihiro}}, \bibinfo {author} {\bibfnamefont {H.}~\bibnamefont {Sakaguchi}},
  \bibinfo {author} {\bibfnamefont {S.}~\bibnamefont {Terashima}}, \bibinfo
  {author} {\bibfnamefont {T.}~\bibnamefont {Uesaka}}, \bibinfo {author}
  {\bibfnamefont {G.}~\bibnamefont {Hagen}}, \bibinfo {author} {\bibfnamefont
  {M.}~\bibnamefont {Itoh}}, \bibinfo {author} {\bibfnamefont {T.}~\bibnamefont
  {Murakami}}, \bibinfo {author} {\bibfnamefont {Y.}~\bibnamefont
  {Nakatsugawa}}, \bibinfo {author} {\bibfnamefont {T.}~\bibnamefont
  {Ohnishi}}, \bibinfo {author} {\bibfnamefont {H.}~\bibnamefont {Sagawa}},
  \bibinfo {author} {\bibfnamefont {H.}~\bibnamefont {Takeda}}, \bibinfo
  {author} {\bibfnamefont {M.}~\bibnamefont {Uchida}}, \bibinfo {author}
  {\bibfnamefont {H.}~\bibnamefont {Yoshida}}, \bibinfo {author} {\bibfnamefont
  {S.}~\bibnamefont {Yoshida}}, \ and\ \bibinfo {author} {\bibfnamefont
  {M.}~\bibnamefont {Yosoi}},\ }\href@noop {} {\  (\bibinfo {year} {2018})},\
  \Eprint {http://arxiv.org/abs/1810.11796} {arXiv:1810.11796 [nucl-ex]}
  \BibitemShut {NoStop}%
\bibitem [{\citenamefont {Press}\ \emph {et~al.}(2007)\citenamefont {Press},
  \citenamefont {Teukolsky}, \citenamefont {Vetterling},\ and\ \citenamefont
  {Flannery}}]{nr}%
  \BibitemOpen
  \bibfield  {author} {\bibinfo {author} {\bibfnamefont {W.~H.}\ \bibnamefont
  {Press}}, \bibinfo {author} {\bibfnamefont {S.~A.}\ \bibnamefont
  {Teukolsky}}, \bibinfo {author} {\bibfnamefont {W.~T.}\ \bibnamefont
  {Vetterling}}, \ and\ \bibinfo {author} {\bibfnamefont {B.~P.}\ \bibnamefont
  {Flannery}},\ }\href@noop {} {\emph {\bibinfo {title} {Numerical recipes}}}\
  (\bibinfo  {publisher} {Cambridge. Unviverstiy Press},\ \bibinfo {year}
  {2007})\ \Eprint
  {http://arxiv.org/abs/http://apps.nrbook.com/empanel/index.html}
  {http://apps.nrbook.com/empanel/index.html} \BibitemShut {NoStop}%
\bibitem [{\citenamefont {Schwierz}\ \emph {et~al.}(2007)\citenamefont
  {Schwierz}, \citenamefont {Wiedenhover},\ and\ \citenamefont
  {Volya}}]{2007Schwierz}%
  \BibitemOpen
  \bibfield  {author} {\bibinfo {author} {\bibfnamefont {N.}~\bibnamefont
  {Schwierz}}, \bibinfo {author} {\bibfnamefont {I.}~\bibnamefont
  {Wiedenhover}}, \ and\ \bibinfo {author} {\bibfnamefont {A.}~\bibnamefont
  {Volya}},\ }\href@noop {} {\  (\bibinfo {year} {2007})},\ \Eprint
  {http://arxiv.org/abs/0709.3525} {arXiv:0709.3525 [nucl-th]} \BibitemShut
  {NoStop}%
\bibitem [{nnd()}]{nndcpag}%
  \BibitemOpen
  \href@noop {} {}\Eprint {http://arxiv.org/abs/www.nndc.gov} {www.nndc.gov}
  \BibitemShut {NoStop}%
\bibitem [{\citenamefont {Vertse}\ \emph {et~al.}(1982)\citenamefont {Vertse},
  \citenamefont {P{\'{a}}l},\ and\ \citenamefont {Balogh}}]{Vertse1982}%
  \BibitemOpen
  \bibfield  {author} {\bibinfo {author} {\bibfnamefont {T.}~\bibnamefont
  {Vertse}}, \bibinfo {author} {\bibfnamefont {K.~F.}\ \bibnamefont
  {P{\'{a}}l}}, \ and\ \bibinfo {author} {\bibfnamefont {Z.}~\bibnamefont
  {Balogh}},\ }\href {\doibase 10.1016/0010-4655(82)90178-3} {\bibfield
  {journal} {\bibinfo  {journal} {Computer Physics Communications}\ }\textbf
  {\bibinfo {volume} {27}},\ \bibinfo {pages} {309} (\bibinfo {year}
  {1982})}\BibitemShut {NoStop}%
\bibitem [{\citenamefont {Ixaru}\ \emph {et~al.}(1995)\citenamefont {Ixaru},
  \citenamefont {Rizea},\ and\ \citenamefont {Vertse}}]{Ixaru1995}%
  \BibitemOpen
  \bibfield  {author} {\bibinfo {author} {\bibfnamefont {L.~G.}\ \bibnamefont
  {Ixaru}}, \bibinfo {author} {\bibfnamefont {M.}~\bibnamefont {Rizea}}, \ and\
  \bibinfo {author} {\bibfnamefont {T.}~\bibnamefont {Vertse}},\ }\href
  {\doibase 10.1016/0010-4655(94)00135-O} {\bibfield  {journal} {\bibinfo
  {journal} {Computer Physics Communications}\ }\textbf {\bibinfo {volume}
  {85}},\ \bibinfo {pages} {217} (\bibinfo {year} {1995})}\BibitemShut
  {NoStop}%
\bibitem [{\citenamefont {Liotta}\ \emph {et~al.}(1996)\citenamefont {Liotta},
  \citenamefont {Maglione}, \citenamefont {Sandulescu},\ and\ \citenamefont
  {Vertse}}]{1996Liotta}%
  \BibitemOpen
  \bibfield  {author} {\bibinfo {author} {\bibfnamefont {R.~J.}\ \bibnamefont
  {Liotta}}, \bibinfo {author} {\bibfnamefont {E.}~\bibnamefont {Maglione}},
  \bibinfo {author} {\bibfnamefont {N.}~\bibnamefont {Sandulescu}}, \ and\
  \bibinfo {author} {\bibfnamefont {T.}~\bibnamefont {Vertse}},\ }\href
  {\doibase https://doi.org/10.1016/0370-2693(95)01415-2} {\ \textbf {\bibinfo
  {volume} {367}},\ \bibinfo {pages} {1 } (\bibinfo {year} {1996})}\BibitemShut
  {NoStop}%
\bibitem [{\citenamefont {Hamamoto}(2012)}]{2012Hamamoto}%
  \BibitemOpen
  \bibfield  {author} {\bibinfo {author} {\bibfnamefont {I.}~\bibnamefont
  {Hamamoto}},\ }\href {\doibase 10.1103/PhysRevC.85.064329} {\bibfield
  {journal} {\bibinfo  {journal} {Physical Review C}\ }\textbf {\bibinfo
  {volume} {85}},\ \bibinfo {pages} {064329} (\bibinfo {year}
  {2012})}\BibitemShut {NoStop}%
\bibitem [{\citenamefont {Wang}\ \emph {et~al.}(2017)\citenamefont {Wang},
  \citenamefont {Audi}, \citenamefont {Kondev}, \citenamefont {Huang},
  \citenamefont {Naimi},\ and\ \citenamefont {Xu}}]{Wang2017}%
  \BibitemOpen
  \bibfield  {author} {\bibinfo {author} {\bibfnamefont {M.}~\bibnamefont
  {Wang}}, \bibinfo {author} {\bibfnamefont {G.}~\bibnamefont {Audi}}, \bibinfo
  {author} {\bibfnamefont {F.~G.}\ \bibnamefont {Kondev}}, \bibinfo {author}
  {\bibfnamefont {W.~J.}\ \bibnamefont {Huang}}, \bibinfo {author}
  {\bibfnamefont {S.}~\bibnamefont {Naimi}}, \ and\ \bibinfo {author}
  {\bibfnamefont {X.}~\bibnamefont {Xu}},\ }\href {\doibase
  10.1088/1674-1137/41/3/030003} {\bibfield  {journal} {\bibinfo  {journal}
  {Chinese Physics C}\ }\textbf {\bibinfo {volume} {41}} (\bibinfo {year}
  {2017}),\ 10.1088/1674-1137/41/3/030003}\BibitemShut {NoStop}%
\bibitem [{\citenamefont {Holt}\ \emph {et~al.}(2019)\citenamefont {Holt},
  \citenamefont {Stroberg}, \citenamefont {Schwenk},\ and\ \citenamefont
  {Simonis}}]{2019Holt}%
  \BibitemOpen
  \bibfield  {author} {\bibinfo {author} {\bibfnamefont {J.~D.}\ \bibnamefont
  {Holt}}, \bibinfo {author} {\bibfnamefont {S.~R.}\ \bibnamefont {Stroberg}},
  \bibinfo {author} {\bibfnamefont {A.}~\bibnamefont {Schwenk}}, \ and\
  \bibinfo {author} {\bibfnamefont {J.}~\bibnamefont {Simonis}},\ }\href@noop
  {} {\  (\bibinfo {year} {2019})},\ \Eprint {http://arxiv.org/abs/1905.10475}
  {arXiv:1905.10475 [nucl-th]} \BibitemShut {NoStop}%
\bibitem [{\citenamefont {Becchetti}\ and\ \citenamefont
  {Greenlees}(1969)}]{PhysRev.182.1190}%
  \BibitemOpen
  \bibfield  {author} {\bibinfo {author} {\bibfnamefont {F.~D.}\ \bibnamefont
  {Becchetti}}\ and\ \bibinfo {author} {\bibfnamefont {G.~W.}\ \bibnamefont
  {Greenlees}},\ }\href {\doibase 10.1103/PhysRev.182.1190} {\bibfield
  {journal} {\bibinfo  {journal} {Phys. Rev.}\ }\textbf {\bibinfo {volume}
  {182}},\ \bibinfo {pages} {1190} (\bibinfo {year} {1969})}\BibitemShut
  {NoStop}%
\bibitem [{\citenamefont {Leistenschneider}\ \emph {et~al.}(2018)\citenamefont
  {Leistenschneider}, \citenamefont {Reiter}, \citenamefont {Ayet
  San~Andr\'es}, \citenamefont {Kootte}, \citenamefont {Holt}, \citenamefont
  {Navr\'atil}, \citenamefont {Babcock}, \citenamefont {Barbieri},
  \citenamefont {Barquest}, \citenamefont {Bergmann}, \citenamefont {Bollig},
  \citenamefont {Brunner}, \citenamefont {Dunling}, \citenamefont {Finlay},
  \citenamefont {Geissel} \emph {et~al.}}]{2018Leistenschneider}%
  \BibitemOpen
  \bibfield  {author} {\bibinfo {author} {\bibfnamefont {E.}~\bibnamefont
  {Leistenschneider}}, \bibinfo {author} {\bibfnamefont {M.~P.}\ \bibnamefont
  {Reiter}}, \bibinfo {author} {\bibfnamefont {S.}~\bibnamefont {Ayet
  San~Andr\'es}}, \bibinfo {author} {\bibfnamefont {B.}~\bibnamefont {Kootte}},
  \bibinfo {author} {\bibfnamefont {J.~D.}\ \bibnamefont {Holt}}, \bibinfo
  {author} {\bibfnamefont {P.}~\bibnamefont {Navr\'atil}}, \bibinfo {author}
  {\bibfnamefont {C.}~\bibnamefont {Babcock}}, \bibinfo {author} {\bibfnamefont
  {C.}~\bibnamefont {Barbieri}}, \bibinfo {author} {\bibfnamefont {B.~R.}\
  \bibnamefont {Barquest}}, \bibinfo {author} {\bibfnamefont {J.}~\bibnamefont
  {Bergmann}}, \bibinfo {author} {\bibfnamefont {J.}~\bibnamefont {Bollig}},
  \bibinfo {author} {\bibfnamefont {T.}~\bibnamefont {Brunner}}, \bibinfo
  {author} {\bibfnamefont {E.}~\bibnamefont {Dunling}}, \bibinfo {author}
  {\bibfnamefont {A.}~\bibnamefont {Finlay}}, \bibinfo {author} {\bibfnamefont
  {H.}~\bibnamefont {Geissel}},  \emph {et~al.},\ }\href {\doibase
  10.1103/PhysRevLett.120.062503} {\bibfield  {journal} {\bibinfo  {journal}
  {Physical Review Letters}\ }\textbf {\bibinfo {volume} {120}},\ \bibinfo
  {pages} {062502} (\bibinfo {year} {2018})}\BibitemShut {NoStop}%
\bibitem [{\citenamefont {Bardeen}\ \emph {et~al.}(1957)\citenamefont
  {Bardeen}, \citenamefont {Cooper},\ and\ \citenamefont
  {Schrieffer}}]{1957Bcs}%
  \BibitemOpen
  \bibfield  {author} {\bibinfo {author} {\bibfnamefont {J.}~\bibnamefont
  {Bardeen}}, \bibinfo {author} {\bibfnamefont {L.~N.}\ \bibnamefont {Cooper}},
  \ and\ \bibinfo {author} {\bibfnamefont {J.~R.}\ \bibnamefont {Schrieffer}},\
  }\href {\doibase 10.1103/PhysRev.108.1175} {\bibfield  {journal} {\bibinfo
  {journal} {Phys. Rev.}\ }\textbf {\bibinfo {volume} {108}},\ \bibinfo {pages}
  {1175} (\bibinfo {year} {1957})}\BibitemShut {NoStop}%
\bibitem [{\citenamefont {Changizi}\ \emph {et~al.}(2015)\citenamefont
  {Changizi}, \citenamefont {Qi},\ and\ \citenamefont {Wyss}}]{Changizi2015}%
  \BibitemOpen
  \bibfield  {author} {\bibinfo {author} {\bibfnamefont {S.~A.}\ \bibnamefont
  {Changizi}}, \bibinfo {author} {\bibfnamefont {C.}~\bibnamefont {Qi}}, \ and\
  \bibinfo {author} {\bibfnamefont {R.}~\bibnamefont {Wyss}},\ }\href {\doibase
  10.1016/j.nuclphysa.2015.04.010} {\bibfield  {journal} {\bibinfo  {journal}
  {Nuclear Physics A}\ }\textbf {\bibinfo {volume} {940}},\ \bibinfo {pages}
  {210} (\bibinfo {year} {2015})}\BibitemShut {NoStop}%
\bibitem [{\citenamefont {Virtanen}\ \emph {et~al.}(2020)\citenamefont
  {Virtanen}, \citenamefont {Gommers}, \citenamefont {Oliphant}, \citenamefont
  {Haberland}, \citenamefont {Reddy}, \citenamefont {Cournapeau}, \citenamefont
  {Burovski}, \citenamefont {Peterson}, \citenamefont {Weckesser},
  \citenamefont {Bright}, \citenamefont {van~der Walt}, \citenamefont {Brett},
  \citenamefont {Wilson}, \citenamefont {Millman}, \citenamefont {Mayorov}
  \emph {et~al.}}]{root}%
  \BibitemOpen
  \bibfield  {author} {\bibinfo {author} {\bibfnamefont {P.}~\bibnamefont
  {Virtanen}}, \bibinfo {author} {\bibfnamefont {R.}~\bibnamefont {Gommers}},
  \bibinfo {author} {\bibfnamefont {T.~E.}\ \bibnamefont {Oliphant}}, \bibinfo
  {author} {\bibfnamefont {M.}~\bibnamefont {Haberland}}, \bibinfo {author}
  {\bibfnamefont {T.}~\bibnamefont {Reddy}}, \bibinfo {author} {\bibfnamefont
  {D.}~\bibnamefont {Cournapeau}}, \bibinfo {author} {\bibfnamefont
  {E.}~\bibnamefont {Burovski}}, \bibinfo {author} {\bibfnamefont
  {P.}~\bibnamefont {Peterson}}, \bibinfo {author} {\bibfnamefont
  {W.}~\bibnamefont {Weckesser}}, \bibinfo {author} {\bibfnamefont
  {J.}~\bibnamefont {Bright}}, \bibinfo {author} {\bibfnamefont {S.~J.}\
  \bibnamefont {van~der Walt}}, \bibinfo {author} {\bibfnamefont
  {M.}~\bibnamefont {Brett}}, \bibinfo {author} {\bibfnamefont
  {J.}~\bibnamefont {Wilson}}, \bibinfo {author} {\bibfnamefont {K.~J.}\
  \bibnamefont {Millman}}, \bibinfo {author} {\bibfnamefont {N.}~\bibnamefont
  {Mayorov}},  \emph {et~al.},\ }\href {\doibase 10.1038/s41592-019-0686-2}
  {\bibfield  {journal} {\bibinfo  {journal} {Nature Methods}\ }\textbf
  {\bibinfo {volume} {17}},\ \bibinfo {pages} {261} (\bibinfo {year}
  {2020})}\BibitemShut {NoStop}%
\end{thebibliography}

\end{document}